\documentclass[twocolumn,preprintnumbers,amsmath,amssymb]{revtex4}

\usepackage{graphicx}
\usepackage{dcolumn}
\usepackage{bm}

\DeclareMathOperator{\erf}{erf}
\DeclareMathOperator{\erfc}{erfc}
\DeclareMathOperator{\erfgau}{erfgau}
\newcommand{\Psimu}{\ensuremath{\Psi^{\mu}}}
\newcommand{\bra}[1]{\ensuremath{\langle #1 \vert}}
\newcommand{\ket}[1]{\ensuremath{\vert #1  \rangle}}

\renewcommand{\b}[1]{\ensuremath{\mathbf{#1}}}

\begin{document}

\title{Long-range/short-range separation of the electron-electron interaction in density functional theory}

\author{Julien Toulouse}
\author{Fran\c{c}ois Colonna}
\author{Andreas Savin}
 \email{savin@lct.jussieu.fr}
\affiliation{
Laboratoire de Chimie Th\'eorique, CNRS et Universit\'e Pierre et Marie Curie,\\
4 place Jussieu, 75252 Paris, France
}

\date{\today}

\begin{abstract}
By splitting the Coulomb interaction into long-range and short-range components, we decompose the energy of a quantum electronic system into long-range and short-range contributions. We show that the long-range part of the energy can be efficiently calculated by traditional wave function methods, while the short-range part can be handled by a density functional. The analysis of this functional with respect to the range of the associated interaction reveals that, in the limit of a very short-range interaction, the short-range exchange-correlation energy can be expressed as a simple local functional of the on-top pair density and its first derivatives. This provides an explanation for the accuracy of the local density approximation (LDA) for the short-range functional. Moreover, this analysis leads also to new simple approximations for the short-range exchange and correlation energies improving the LDA.
\end{abstract}

\maketitle

\section{Introduction}
\label{sec:intro}

In the Kohn-Sham (KS) approach~\cite{KohSha-PR-65} of density functional theory (DFT)~\cite{HohKoh-PR-64} of inhomogeneous electronic systems, the central quantity is the unknown exchange-correlation energy functional $E_{xc}[n]$ which encompasses all the many-body effects. The vast majority of approximations for this functional are based on the original local density approximation (LDA)~\cite{HohKoh-PR-64}, an approximation that turned out to be more accurate and reliable than expected and rather difficult to improve in a systematic way~\cite{JonGun-RMP-89}.

Actually, it has been realized for long, with the wave-vector analysis of $E_{xc}[n]$ by Langreth and Perdew~\cite{LanPer-PRb-77}, that the LDA describes accurately (but not exactly in general~\cite{BurPerLan-PRL-94}) short wavelength density fluctuations, but is inadequate for long wavelength fluctuations. A dual analysis in real space of the exchange-correlation energy (see, e.g., Ref.~\onlinecite{BurPer-IJQC-95}) leads to the same conclusion that the LDA is accurate at small interelectronic distances but fails at large distances. This observation lead to the development of the first gradient corrected functionals~\cite{LanMeh-PRL-81,LanMeh-PRB-83,Per-PRL-85,PerWan-PRB-86,Per-PRB-86,Per-INC-91} with the basic objective to cure the wrong long-range contribution to the exchange-correlation energy of the LDA.

However, describing accurately the non-local correlation effects arising from the long-range character of the Coulomb interaction by (semi)local density functional approximations still seems out of reach. This idea in mind, it is has been proposed to use a density functional approximation only for the short-range part of the electronic energy, and treating the long-range part by more appropriate many-body methods~\cite{KohHan-JJJ-XX,KohMeiMak-PRL-98,StoSav-INC-85,Sav-INC-96a,Sav-INC-96}. This approach was somehow inspired from early calculations of the correlation energy of the uniform electron gas based on a separate treatment of long-range and short-range contributions (see, e.g., Refs.~\onlinecite{NozPin-PR-58,Rai-BOOK-72,GroRunHei-BOOK-91}).

For atoms and molecules, this approach leads to a rigorous method for combining traditional \textit{ab initio} wave function calculations with DFT~\cite{LeiStoWerSav-CPL-97,PolSavLeiSto-JCP-02}, and we recall now the formalism. The Coulomb electron-electron interaction is decomposed as
\begin{equation}
\frac{1}{r} = v_{ee}^{\mu}(r) + \bar{v}_{ee}^{\mu}(r),
\end{equation}
where $v_{ee}^{\mu}(r)$ is a long-range interaction, $\bar{v}_{ee}^{\mu}(r)$ is the complement short-range interaction and $\mu$ is a parameter controlling the separation. The universal functional~\cite{Lev-PNAS-79} $F[n]=\min_{\Psi\to n} \bra{\Psi} \hat{T}+\hat{V}_{ee} \ket{\Psi}$, where $\hat{T}$ is the kinetic energy operator and $\hat{V}_{ee}=\sum_{i<j} 1/r_{ij}$ is the Coulomb interaction operator, can then be decomposed as
\begin{equation}
F[n] = F^{\mu}[n] + \bar{F}^{\mu}[n],
\end{equation}
where $F^{\mu}[n]$ is the universal functional corresponding to the long-range interaction $\hat{V}_{ee}^{\mu}=\sum_{i<j}v_{ee}^{\mu}(r_{ij})$
\begin{equation}
F^{\mu}[n]= \min_{\Psi\to n} \bra{\Psi}\hat{T} + \hat{V}_{ee}^{\mu}\ket{\Psi}, 
\label{Fmu}
\end{equation}
and $\bar{F}^{\mu}[n]=F[n] - F^{\mu}[n]$ is by definition the complement (short-range) part.

The exact ground-state energy of an electronic system in the external local nuclei-electron potential $v_{ne}(\b{r})$ can be written using this short-range functional $\bar{F}^{\mu}[n]$ via application of the variational principle
\begin{eqnarray}
 E &=& \min_{n} \Bigl\{ F^{\mu}[n] + \bar{F}^{\mu}[n] + \int n(\b{r}) v_{ne}(\b{r}) d\b{r} \Bigl\}
\nonumber\\
   &=& \min_{\Psi} \Bigl\{ \bra{\Psi} \hat{T}+\hat{V}_{ee}^{\mu} \ket{\Psi} + \bar{F}^{\mu}[n_{\Psi}] + \int n_{\Psi}(\b{r}) v_{ne}(\b{r}) d\b{r} \Bigl\}
\nonumber\\
 &=& \bra{\Psimu} \hat{T}+\hat{V}_{ee}^{\mu} \ket{\Psimu} + \bar{F}^{\mu}[n_{\Psimu}] + \int n_{\Psimu}(\b{r}) v_{ne}(\b{r}) d\b{r}.
\nonumber\\
\label{E}
\end{eqnarray}
In this equation, $\Psimu$ is given by the Euler-Lagrange equation
\begin{equation}
 \hat{H}^{\mu} \ket{\Psimu} = E^{\mu} \ket{\Psimu},
\label{HmuPsimu}
\end{equation}
where $\hat{H}^{\mu}=\hat{T}+\hat{V}_{ee}^{\mu}+\hat{V}^{\mu}$. Thus, $\Psimu$ is the ground-state multi-determinantal wave function of a partially interacting system with interaction $\hat{V}_{ee}^{\mu}$ and external local potential $\hat{V}^{\mu}=\sum_{i}v^{\mu}(\b{r}_i)$ where $v^{\mu}(\b{r})=v_{ne}(\b{r}) + \delta \bar{F}^{\mu}[n_{\Psimu}]/\delta n(\b{r})$. By virtue of the Hohenberg-Kohn theorem~\cite{HohKoh-PR-64}, $v^{\mu}(\b{r})$ is the unique potential (up to an additive constant) which insures that the ground-state density of this fictitious system is identical to the ground-state density of the physical system $n_{\Psimu}=n$.

The short-range functional $\bar{F}^{\mu}[n]$ is further decomposed as
\begin{equation}
\bar{F}^{\mu}[n] = \bar{U}^{\mu}[n] + \bar{E}_{xc}^{\mu}[n],
\end{equation}
where $\bar{U}^{\mu}[n]=1/2 \iint n(\b{r}_1)n(\b{r}_1)\bar{v}_{ee}^{\mu}(r_{12}) d\b{r}_1 d\b{r}_2$ is the complement short-range Hartree energy functional and $\bar{E}_{xc}^{\mu}[n]$ is the unknown complement short-range exchange-correlation energy functional. Once an approximation is chosen for $\bar{E}_{xc}^{\mu}[n]$, the wave function $\Psimu$ can be computed by solving self-consistently Eq.~(\ref{HmuPsimu}) using \textit{ab initio} wave function methods like configuration interaction (CI) or multi-configurational self-consistency field (MCSCF) and the total energy is calculated according to the last line of Eq.~(\ref{E}). Notice also that it is possible to separate the functional $\bar{E}_{xc}^{\mu}[n]$ into short-range exchange and correlation contributions, $\bar{E}_{xc}^{\mu}[n]=\bar{E}_{x}^{\mu}[n]+\bar{E}_{c}^{\mu}[n]$, with
\begin{equation}
\label{Ex}
\bar{E}_{x}^{\mu}[n] = \bra{\Phi} \hat{\bar{V}}_{ee}^{\mu} \ket{\Phi} - \bar{U}^{\mu}[n],
\end{equation}
and
\begin{eqnarray}
\bar{E}_{c}^{\mu}[n] = \bar{F}^{\mu}[n] - \bra{\Phi} \hat{\bar{V}}_{ee}^{\mu} \ket{\Phi}
\label{Ec}
\end{eqnarray}
where $\Phi$ is the KS determinant and $\hat{\bar{V}}_{ee}^{\mu}=\hat{V}_{ee} - \hat{V}_{ee}^{\mu}$ is the complement short-range interaction operator. 

A simple approximation for $\bar{E}_{xc}^{\mu}[n]$ is the LDA associated to the modified interaction~\cite{Sav-INC-96,TouSavFla-IJQC-XX}
\begin{equation}
\bar{E}^{\mu,\text{LDA}}_{xc}[n]= \int n(\b{r}) \bar{\varepsilon}^{\mu,\text{unif}}_{xc}(n(\b{r})) d\b{r},
\label{LDA}
\end{equation}
where $\bar{\varepsilon}^{\mu,\text{unif}}_{xc}(n)$ is the complement short-range exchange-correlation energy per particle obtained by difference from the exchange-correlation energies per particle of the uniform electron gas with the standard Coulomb interaction, $\varepsilon_{xc}^{\text{unif}}(n)$, and with the long-range interaction $v_{ee}^{\mu}$, $\varepsilon^{\mu,\text{unif}}_{xc}(n)$,
\begin{equation}
\bar{\varepsilon}^{\mu,\text{unif}}_{xc}(n) = \varepsilon_{xc}^{\text{unif}}(n) - \varepsilon^{\mu,\text{unif}}_{xc}(n).
\end{equation}
The method has been implemented at an experimental level into the quantum chemistry package Molpro, allowing a combination of CI-type wave function and DFT calculations~\cite{LeiStoWerSav-CPL-97}. Recently, the method has also been efficiently implemented into the quantum chemistry program Dalton, performing the coupling of MCSCF wave function calculations with DFT~\cite{PedJen-JJJ-XX}. For a reasonable long-range/short-range separation and using only the LDA for $\bar{E}_{xc}^{\mu}[n]$, the approach already yields good results for total energies of atomic and molecular systems~\cite{LeiStoWerSav-CPL-97,PolSavLeiSto-JCP-02,PedJen-JJJ-XX}. The purpose of this paper is to analyzed in further details the two separate needed approximations of the method: the wave function calculation of Eq.~(\ref{HmuPsimu}) and the short-range exchange-correlation functional $\bar{E}_{xc}^{\mu}[n]$. Accurate calculations have been performed on a few small atomic systems to assess the approximations. Concerning the wave function part, we show that the modification of the interaction enables to decrease the effort, or alternatively increase the accuracy, of a CI-type calculation. Concerning the functional part, we gain more insights into the short-range functional and explain the performance of the LDA by studying the behavior of $\bar{E}_{xc}^{\mu}[n]$ with respect to the interaction parameter $\mu$. In particular, we show that, when the interaction $\bar{v}^{\mu}_{ee}$ is short ranged enough, $\bar{E}_{xc}^{\mu}[n]$ can be expressed as a simple local functional of the on-top pair density and its first derivatives, explaining the accuracy of (semi)local functional approximations. This analysis also enables us to propose new approximations for $\bar{E}_{xc}^{\mu}[n]$ which correct the LDA in the domain of $\mu$ where it fails.

The paper is organized as follows. In Sec.~\ref{sec:lrsrseparation}, we give two possible choices for the long-range interaction $v_{ee}^{\mu}(r)$. In Sec.~\ref{sec:accurate}, technical details concerning the calculations made on atomic systems are given. In Sec.~\ref{sec:wavefunction}, the impact of the modification of the interaction on the performance of wave function calculation is investigated. In Sec.~\ref{sec:srxc}, we study the behavior of the corresponding short-range exchange and correlation functionals with respect to the interaction parameter $\mu$. In Sec.~\ref{sec:discussion}, we test the obtained exact behaviors and compared them with the LDA. In Sec~\ref{sec:approx}, interpolations for the short-range exchange and correlation functionals with respect to $\mu$ are proposed and tested. Finally, Sec~\ref{sec:conclusion} draws a conclusion of this work.

Atomic units (a.u.) will be used throughout this work.

\section{Long-range/short-range separation}
   \label{sec:lrsrseparation}

In a number of previous works~\cite{Sav-INC-96,LeiStoWerSav-CPL-97,PolSavLeiSto-JCP-02,PolColLeiStoWerSav-IJQC-03,SavColPol-IJQC-03}, a splitting of the Coulomb interaction based on the error function has been studied to describe the long-range part of the electron-electron interaction
\begin{equation}
\label{veeerf}
v_{ee,\erf}^{\mu}(r)=\frac{\erf(\mu r)}{r}.
\end{equation}
This interaction, referred to as the erf interaction, has also been used by other authors in DFT for various purposes~\cite{GilAda-CPL-96,GilAdaPop-MP-96,IikTsuYanHir-JCP-01,KamTsuHir-JCP-02,TawTsuYanYanHir-JCP-04,YanTewHan-CPL-04,HeyScuErn-JCP-03,HeyScu-JCP-04}. In this work, we introduce another interaction achieving a sharper separation of long-range and short-range interactions by subtracting a Gaussian function from the erf interaction
\begin{equation}
\label{veeerfgau}
v_{ee,\erfgau}^{\mu}(r)=\frac{\erf(\mu r)}{r} - \frac{2\mu}{\sqrt{\pi}} e^{-\frac{1}{3}\mu^2 r^2},
\end{equation}
where the coefficient and the exponent of the Gaussian have been chosen so that $v_{ee,\erfgau}^{\mu}(r)$ and its derivative with respect to $r$ vanish at $r=0$. This modified interaction is referred to as the erfgau interaction. Notice that this partition of Coulomb interaction has already been proposed by Gill and Adamson~\cite{GilAda-CPL-96}. In another context, Prendergast \textit{et al.}~\cite{PreNolFilFahGre-JCP-01} have also used a similar form of long-range interaction. For the sake of completeness, we finally note that another possible form of modified electron-electron interaction based on the Yukawa potential has also been investigated in the past~\cite{KohHan-JJJ-XX,StoSav-INC-85,Sav-INC-96a,SavFla-IJQC-95,KohMeiMak-PRL-98,ArmMat-PRB-03}. 

Both interactions~(\ref{veeerf}) and~(\ref{veeerfgau}) enable to define a generalized adiabatic connection~\cite{Yan-JCP-98} between the non-interacting KS system at $\mu=0$ where the interaction vanishes $v_{ee}^{\mu=0}(r)=0$, and the physical system at $\mu \to \infty$ where the full Coulomb interaction is recovered $v_{ee}^{\mu \to \infty}(r)=1/r$. 

\begin{figure}
\includegraphics[scale=0.75]{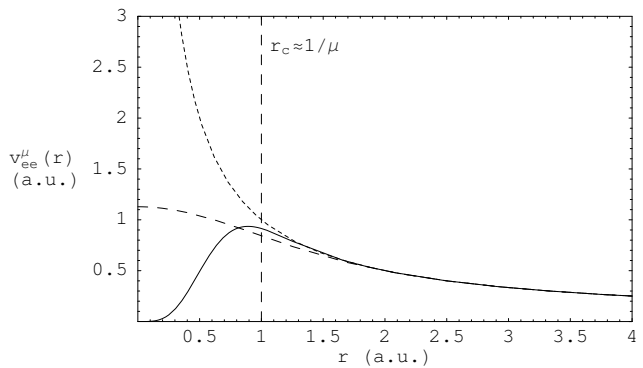}
\caption{Coulomb interaction $1/r$ (dotted curve), erf interaction (dashed curve) and erfgau interaction (solid curve) for $\mu=1$. In order to compare the two modified interactions, a scale factor is applied on the interaction parameter of the erfgau interaction: $\mu \to c \mu$  with $c= 3.375$ (see text). It is possible to define a common cut-off radius $r_c \approx 1/\mu$.
}
\label{fig:vee-erferfgau}
\end{figure}

Notice that, although we have chosen the same notation for convenience, the parameter $\mu$ in Eq.~(\ref{veeerf}) is \textit{a priori} independent of that of Eq.~(\ref{veeerfgau}). Actually, in all the plots of the paper involving the erfgau interaction, we apply a scale factor to the interaction parameter of the erfgau interaction: $\mu \to c \mu$. The constant $c$ is chosen so as to have the same leading term in the distributional asymptotic expansion of the two interactions when $\mu \to \infty$ (see Appendix~\ref{app:asymptoticexpansion}) which leads to the value $c = \left( 1+6\sqrt{3}\right)^{1/2} \approx 3.375$. This also insures that the leading term in the asymptotic expansion of short-range Hartree, exchange and correlation energies for large $\mu$ is the same for the two interactions. In Fig.~\ref{fig:vee-erferfgau}, the erf and erfgau interactions are compared along with the Coulomb interaction. The scale factor on the parameter of the erfgau interaction enables to define a common ``cut-off radius'' giving the range of the interaction and defined by the inverse of the interaction parameter $r_c \approx 1/\mu$. For interelectronic distances larger than $r_c$, the two modified interactions reproduce the long-range Coulomb tail. Notice that short-range interactions are better removed with the erfgau interaction.

We finally note that, for Gaussian basis set calculations, the evaluation of the two-electron integrals corresponding to the erf or erfgau interaction requires only simple modifications of standard algorithms for Coulomb integrals (see Appendix~\ref{app:integrals}).

\section{Details on accurate calculations}
\label{sec:accurate}

We explain rapidly how the accurate data for atomic systems presented in this work have been obtained.

For each system, accurate potentials $v^{\mu}(\b{r})$ needed in Eq.~(\ref{HmuPsimu}) and corresponding value of the universal long-range functional $F^{\mu}$ are computed using the Lieb's Legendre transform formulation of the universal functional~\cite{Lie-IJQC-83}
\begin{equation}
\label{FLieb}
F^{\mu}[n]= \max_{\tilde{v}^{\mu}} \Bigl\{ E^{\mu}[\tilde{v}^{\mu}] - \int n(\b{r}) \tilde{v}^{\mu}(\b{r}) d\b{r} \Bigl\},
\end{equation}
where $E^{\mu}[\tilde{v}^{\mu}]$ is the ground-state energy of the Hamiltonian $\hat{T} + \hat{V}_{ee}^{\mu} + \sum_i \tilde{v}^{\mu}(\b{r}_i)$. If $n$ is chosen to be the physical density of the system, the maximum is reached for the desired $v^{\mu}$ and $F^{\mu}$. In practice, an accurate density $n$ is computed by multi-reference CI with single and double excitations (MRCISD)~\cite{KnoWer-CPL-88,WerKno-JCP-88} and the potential to optimize $\tilde{v}^{\mu}(r)$ is expanded as
\begin{equation}
\tilde{v}^{\mu}(r) = \sum_{i=1}^{n} c_i r^{p_i} e^{\gamma_i r^2} + \frac{C}{r},
\end{equation}
where $c_i$ are the optimized coefficients, $p_i$ are some fixed integers ($-1$ or $2$), $\gamma_i$ are fixed exponents chosen so as to form an even-tempered basis set (typically, $\gamma_i \in [10^{-3}, 5.10^4]$), and $C$ is a constant enforces the correct asymptotic behavior for $r \to \infty$. For the Kohn-Sham case
($\mu=0$), $C=-Z+N-1$ whereas for finite $\mu$, $C=-Z$ ($N$ and $Z$ are the electron number and nuclear charge, respectively). 

The maximization of Eq.~(\ref{FLieb}) is carried out with the Simplex method~\cite{PreTeuVetFla-BOOK-92}. For a given potential, $E^{\mu}[\tilde{v}^{\mu}]$ is computed at MRCISD level using Molpro~\cite{Molproshort-PROG-02} with modified two-electron integrals (see Appendix~\ref{app:integrals}). Beside the asymptotic behavior for $r \to \infty$, $v^{\mu}(r) \sim C/r$, the behavior of the potential at the nucleus $r=0$, $v^{\mu}(r) \sim -Z/r$, is also imposed during the optimization. Large one-electron even-tempered Gaussian basis sets are used for all systems (see Refs.~\onlinecite{ColSav-JCP-99} and~\onlinecite{PolColLeiStoWerSav-IJQC-03} for more details). 

The standard universal functional $F$ and the KS potential $v_{KS}$ (and thus the KS determinant $\Phi$) are obtained as a special case for $\mu=0$. The complement short-range functional $\bar{F}^{\mu}= F - F^{\mu}$ can then be deduced and the short-range exchange and correlation energies, $\bar{E}_{x}^{\mu}$ and $\bar{E}_{c}^{\mu}$, are obtained from Eqs.~(\ref{Ex}) and~(\ref{Ec}).

\section{Performance of the approximations for the wave function calculation}
\label{sec:wavefunction}

We now investigate the effect of the modification of the electron-electron interaction on the wave function part of the calculation. For this purpose, we evaluate the efficiency of approximate resolutions of Eq.~(\ref{HmuPsimu}) as follows.

We first construct, for each $\mu$, the Hamiltonian $\hat{H}^{\mu}$ using an accurate potential $v^{\mu}$ and compute accurately its ground-state energy, $E^{\mu} = \bra{\Psimu} \hat{H}^{\mu} \ket{\Psimu}$, at the MRCISD level. We then compute various approximate ground-state energies, $E^{\mu}_{S} = \bra{\Psi^{\mu}_{S}} \hat{H}^{\mu} \ket{\Psi^{\mu}_{S}}$, by using approximate CI-type wave functions $\Psi^{\mu}_{S}$ expanded into linear combinations of all the few Slater determinants generated from small orbital spaces $S$. The orbitals used are the natural orbitals of the Coulombic system calculated at the MRCISD level. The accuracy of the approximation for $\Psi^{\mu}_{S}$ can be assessed by looking at the difference between $E^{\mu}_{S}$ and $E^{\mu}$ 
\begin{equation}
\Delta E^{\mu}_{S}= E^{\mu}_{S} - E^{\mu}.
\end{equation}

\begin{figure}
\includegraphics[scale=0.75]{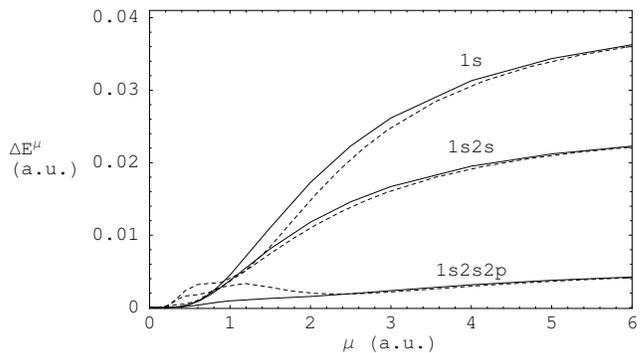}
\caption{Ground-state energy differences $\Delta E^{\mu}_{S}= \bra{\Psi^{\mu}_{S}} \hat{H}^{\mu} \ket{\Psi^{\mu}_{S}} - \bra{\Psimu} \hat{H}^{\mu} \ket{\Psimu}$ where $\Psimu$ is an accurate wave function and $\Psi^{\mu}_{S}$ are approximate wave functions generated from small orbital spaces $S=1s$, $S=1s2s$ and $S=1s2s2p$ along the erf (full curves) and erfgau (dashed curves) adiabatic connections for He.
}
\label{fig:config-he-erferfgau}
\end{figure}

The differences $\Delta E^{\mu}_{S}$ are plotted along the erf and erfgau adiabatic connections in Fig.~\ref{fig:config-he-erferfgau} for the He atom with the orbital spaces $S=1s$, $S=1s2s$ and $S=1s2s2p$. One sees that, in the Coulombic limit $\mu \to \infty$, the reduction of the orbital space leads to important errors in the energy. When $\mu$ is decreased, i.e. when the interaction is reduced, the errors due to limited orbital spaces get smaller and smaller. For instance, at $\mu=1$, using only the single-determinant wave function $\Psimu_{1s}$, leads to an error $\Delta E^{\mu}_{1s}$ of less than $0.005$ Hartree. The erfgau interaction generally gives smaller errors than the erf interaction, except near $\mu=0$ where the non-monotonicity of the erfgau interaction leads to peculiar behaviors~\cite{TouSavFla-IJQC-XX}. Anyway, we will not use the erfgau interaction in this region of very small $\mu$.

\begin{figure}
\includegraphics[scale=0.75]{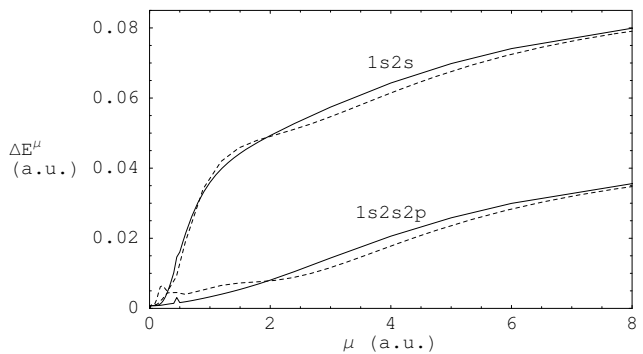}
\caption{
Ground-state energy differences $\Delta E^{\mu}_{S}= \bra{\Psi^{\mu}_{S}} \hat{H}^{\mu} \ket{\Psi^{\mu}_{S}} - \bra{\Psimu} \hat{H}^{\mu} \ket{\Psimu}$ where $\Psimu$ is an accurate wave function and $\Psi^{\mu}_{S}$ are approximate wave functions generated from small orbital spaces $S=1s2s$ and $S=1s2s2p$ along the erf (full curves) and erfgau (dashed curves) adiabatic connections for Be.
}
\label{fig:config-be-erferfgau}
\end{figure}

The case of the Be atom with the orbital spaces $S=1s2s$ and $S=1s2s2p$ is reported in Fig.~\ref{fig:config-be-erferfgau}. Because of the near-degeneracy of the $2s$ and $2p$ levels, the inclusion of $2p$ configurations in the wave function is important, quite independently of the electron-electron interaction. Indeed, the difference $E^{\mu}_{1s2s} - E^{\mu}_{1s2s2p}$  remains large for almost all $\mu$'s. On the contrary, the error of the calculation where the $2p$ orbitals are included, $\Delta E^{\mu}_{1s2s2p}$, quickly falls off when $\mu$ is decreased. Again, for $\mu=1$ for instance, the error $\Delta E^{\mu}_{1s2s2p}$ given by the few-determinant CI-type wave function $\Psimu_{1s2s2p}$ is less than $0.005$ Hartree. 

Therefore, the modification of the interaction enables to increase the accuracy of CI-type wave function calculations, or equivalently for a fixed target accuracy, decrease the effort of the calculation by reducing the orbital space. The crucial point for this effect to appear is the reduction of the electron-electron interaction compared to the Coulomb interaction and not really the long-range character of the modified interaction.

\section{Behavior of short-range exchange and correlation functionals with respect to the interaction parameter}
   \label{sec:srxc}

At $\mu=0$, the short-range exchange-correlation functional reduces to the standard exchange-correlation functional of the KS scheme, $\bar{E}_{xc}^{\mu=0}=E_{xc}$, and in the limit $\mu \to \infty$, the short-range functional vanishes, $\bar{E}_{xc}^{\mu \to \infty}=0$. Near these two limits, the study of the behavior of $\bar{E}_{x}^{\mu}$ and $\bar{E}_{c}^{\mu}$ with respect to $\mu$ constitutes an analysis of the exchange and correlation functionals in term of the range of the interaction. Indeed, the behavior at $\mu \to 0$ tells us how the KS exchange-correlation functional responds when very long-range interactions are removed from it, while the asymptotic expansion for $\mu \to \infty$ gives the exchange-correlation functional associated to very short-range interactions. To release the text from mathematical details, the full derivation of the expansions are given in the Appendices.

      \subsection{Exchange functional for small $\mu$}
In Appendix~\ref{app:expansionsmallmu}, we show that the short-range exchange energy has the following formal expansion around $\mu=0$
\begin{eqnarray}
\bar{E}_{x}^{\mu} &=& E_{x} - \frac{1}{\sqrt{\pi}} \sum_{n=0}^{\infty} \frac{(-1)^n a_n}{n!} \mu^{2n+1} \nonumber\\
 & & \times \iint n_{2,x}(\b{r}_1,\b{r}_2) r_{12}^{2n} d\b{r}_1 d\b{r}_2,
\label{Exsrmu0text}
\end{eqnarray}
where $E_{x}$ is the usual KS exchange energy, $n_{2,x}(\b{r}_1,\b{r}_2)$ is the exchange contribution to the pair density, and $a_n$ are coefficients depending of the interaction chosen and defined after Eq.~(\ref{veemclaurin}). More specifically, for the erf interaction, this expansion writes
\begin{equation}
\label{Exsrmu0erf}
\bar{E}_{x,\erf}^{\mu} = E_{x} + \frac{\mu}{\sqrt{\pi}} N + \frac{\mu^3}{3 \sqrt{\pi}} \iint n_{2,x}(\b{r}_1,\b{r}_2) r_{12}^{2} d\b{r}_1 d\b{r}_2  + \cdots,
\end{equation}
where the term linear in $\mu$ comes from the normalization of the exchange hole~\cite{Yan-JCP-98}. For the erfgau interaction, the expansion is
\begin{equation}
\label{Exsrmu0erfgau}
\bar{E}_{x,\erfgau}^{\mu} = E_{x} - \frac{2\mu^5}{45 \sqrt{\pi}} \iint n_{2,x}(\b{r}_1,\b{r}_2) r_{12}^4 d\b{r}_1 d\b{r}_2 + \cdots.
\end{equation}
The exchange energy with the erf interaction varies linearly in $\mu$ near the Kohn-Sham end of the adiabatic connection (see Fig.~\ref{fig:ex-he-erf-lda}). On the contrary, the exchange energy with the erfgau interaction varies very slowly like $\mu^5$ which implies that the corresponding curve is very flat near $\mu=0$ (see Fig.~\ref{fig:ex-he-erfgau-lda}). The latter behavior is not unexpected for a good long-range/short-range separation. Indeed, with the erfgau interaction, when $\mu$ increases near $\mu=0$ only very long-range interaction effects are removed from the functional; they practically do not later the exchange energy of a finite system. With the erf interaction, the long-range/short-range separation is imperfect and thus the short-range part of the interaction is also affected near $\mu=0$ which is responsible for the linear behavior of the exchange energy in this case. This clearly shows that the erfgau interaction realizes a better separation of long-range and short-range interactions than the erf interaction.

      \subsection{Correlation functional for small $\mu$}
The general expansion of short-range correlation energy around $\mu=0$ is derived in Appendix~\ref{app:expansionsmallmu}. It reads
\begin{eqnarray}
\bar{E}_{c}^{\mu} &=& E_c - \frac{1}{\sqrt{\pi}} \sum_{n=1}^{\infty} \sum_{k=1}^{\infty} \frac{(-1)^n (2n+1) a_n}{n! k! (2n+k+1)} \mu^{2n+k+1} 
\nonumber\\
 & & \times \iint \left( \frac{\partial^k n_{2,c}^{\mu}(\b{r}_1,\b{r}_2)}{\partial \mu^k} \right)_{\mu=0} r_{12}^{2n} d\b{r}_1 d\b{r}_2.
\label{Ecsrmu0text}
\end{eqnarray}
where $E_c$ is the usual correlation energy of the KS scheme and $n_{2,c}^{\mu}(\b{r}_1,\b{r}_{2})$ is the correlation pair density with interaction $v_{ee}^{\mu}$.
Actually, several terms of this expansion vanish. For the erf interaction, the expansion writes
\begin{eqnarray}
\bar{E}_{c,\erf}^{\mu} &=& E_{c} + \frac{\mu^6}{36 \sqrt{\pi}} 
\nonumber\\
&& \times \iint \left( \frac{\partial^3 n_{2,c}^{\mu}(\b{r}_1,\b{r}_2)}{\partial \mu^3} \right)_{\mu=0} r_{12}^{2} d\b{r}_1 d\b{r}_2 + \cdots.
\nonumber\\
\label{Ecsrmu0erf}
\end{eqnarray}
Similarly, the expansion for the erfgau interaction is
\begin{eqnarray}
\bar{E}_{c,\erfgau}^{\mu} &=& E_{c} - \frac{\mu^{10}}{5400 \sqrt{\pi}}
\nonumber\\ 
 & & \times \iint \left( \frac{\partial^5 n_{2,c}^{\mu}(\b{r}_1,\b{r}_2)}{\partial \mu^5} \right)_{\mu=0} r_{12}^{4} d\b{r}_1 d\b{r}_2 + \cdots.
\nonumber\\
\label{Ecsrmu0erfgau}
\end{eqnarray}
The correlation energy varies much more slowly than the exchange energy near the KS end of the adiabatic connection. Therefore, the curve of the correlation energy with respect to $\mu$ is very flat around $\mu=0$ (see Figs.~\ref{fig:ec-he-erf-lda} and~\ref{fig:ec-he-erfgau-lda}). Again, this is due to the fact that removing very long-range interactions from the functional has no effect in a finite system. It will be seen in Sec~\ref{sec:approx} that this makes the correlation energy difficult to interpolate near $\mu=0$ from a knowledge of the functional for large $\mu$.

  \subsection{Exchange functional for large $\mu$}
In Appendix~\ref{app:asymptoticexpansion}, we derive the general asymptotic expansion of the short-range exchange energy for $\mu \to \infty$
\begin{eqnarray}
\bar{E}_{x}^{\mu} &=& 2\sqrt{\pi} \sum_{n=0}^{m} \frac{A_{2n}}{(2n)! (2n+2) \mu^{2n+2}}  
\nonumber\\
&& \times \int n_{2,x}^{(2n)}(\b{r},\b{r})  d\b{r} + {\cal O}(\frac{1}{\mu^{2m+3}}),
\label{Exsrmuinftext}
\end{eqnarray}
where $n_{2,x}^{(2n)}(\b{r},\b{r})$ are the on-top exchange pair density and its spherical-averaged (with respect to $\b{r}_{12}$) derivatives. Simple explicit expressions can be given for the first two terms of this expansion. Indeed, the on-top exchange pair density writes
\begin{equation}
n_{2,x}(\b{r},\b{r})=- \sum_{\sigma} n_{\sigma}^2(\b{r}),
\end{equation}
where the summation is over the two spin states $\sigma=\alpha,\beta$ and $n_{\sigma}(\b{r})$ are the spin-densities; its second-derivative can be expressed by~\cite{Bec-IJQC-83}
\begin{eqnarray}
n_{2,x}^{(2)}(\b{r},\b{r}) &=&- \frac{1}{3} \sum_{\sigma} n_{\sigma}(\b{r})
\nonumber\\
&& \times \left( \nabla^2 n_{\sigma}(\b{r}) - 4 \tau_{\sigma}(\b{r}) +\frac{1}{2} \frac{|\nabla n_{\sigma}(\b{r})|^2}{n_{\sigma}(\b{r})}\right),
\nonumber\\
\label{n2x2}
\end{eqnarray}
where $\tau_{\sigma}(\b{r})$ are the KS spin kinetic energy densities expressed in term of the KS spin-orbitals $\phi_{i \sigma}(\b{r})$ by
\begin{equation}
\tau_{\sigma}(\b{r})=\frac{1}{2} \sum_{i=1}^{N_{\sigma}} |\nabla \phi_{i \sigma}(\b{r})|^2.
\end{equation}
where $N_{\sigma}$ is the number of electrons of spin $\sigma$. The Laplacian in Eq.~(\ref{n2x2}) can be eliminated by integration by parts
\begin{equation}
\int n_{\sigma}(\b{r}) \nabla^2 n_{\sigma}(\b{r}) d\b{r} = - \int |\nabla n_{\sigma}(\b{r})|^2  d\b{r},
\end{equation}
which leads for the leading terms of the expansion when $\mu\to\infty$ of the short-range exchange energy
\begin{eqnarray}
\bar{E}_{x}^{\mu} &=& -\frac{\sqrt{\pi} A_0}{\mu^2} \sum_{\sigma} \int n_{\sigma}(\b{r})^2 d\b{r}
\nonumber\\
&& + \frac{\sqrt{\pi} A_2}{12\mu^4} \sum_{\sigma} \int n_{\sigma}(\b{r}) \left(\frac{|\nabla n_{\sigma}(\b{r})|^2}{2n_{\sigma}(\b{r})} +4\tau_{\sigma}(\b{r}) \right) d\b{r}
\nonumber\\
&& + \cdots,
\label{Exsrmuinf2spin}
\end{eqnarray}
with $A_{0,\erf}=\sqrt{\pi}/2$, $A_{0,\erfgau}=(1+6\sqrt{3})A_{0,\erf}$, $A_{2,\erf}=3\sqrt{\pi}/4$ and $A_{2,\erfgau}=(1+36\sqrt{3})A_{2,\erf}$. Note that this last result has already been demonstrated for the case of the erf interaction~\cite{GilAdaPop-MP-96}. In particular, for spin-unpolarized systems, Eq.~(\ref{Exsrmuinf2spin}) becomes
\begin{eqnarray}
\bar{E}_{x}^{\mu} &=& -\frac{\sqrt{\pi} A_0}{2\mu^2} \int n(\b{r})^2 d\b{r}
\nonumber\\
&& + \frac{\sqrt{\pi} A_2}{24\mu^4} \int n(\b{r}) \left(\frac{|\nabla n(\b{r})|^2}{2n(\b{r})} +4\tau(\b{r}) \right) d\b{r} + \cdots.
\nonumber\\
\label{Exsrmuinf2}
\end{eqnarray}
Eq.~(\ref{Exsrmuinf2spin}) or~(\ref{Exsrmuinf2}) shows that the first term in the expansion for large $\mu$ of the short-range exchange energy is an exact local functional of the density (or, in general, of the spin-densities). Therefore, the local density approximation becomes exact in this limit. An alternative view of this result can be achieved in the framework of the wave-vector analysis of the usual KS exchange functional by considering the short wavelength limit~\cite{BurPerLan-PRL-94}. The next term of expansion~(\ref{Exsrmuinf2}) involves the gradient of the density $|\nabla n|$ and the kinetic energy density $\tau$ and is therefore of the meta-GGA type. The following higher-order terms involve of course more and more ingredients constructed from higher-order derivatives of the KS orbitals. Note that instead of considering these spin-dependent quantities one can also directly use the on-top exchange pair density and its derivatives.

A similar asymptotic expansion can be derived for the short-range Hartree energy
\begin{eqnarray}
\bar{U}^{\mu} &=& \frac{\sqrt{\pi} A_0}{\mu^2} \int n(\b{r})^2 d\b{r}
\nonumber\\
&& - \frac{\sqrt{\pi} A_2}{12\mu^4} \int |\nabla n(\b{r})|^2 d\b{r} + \cdots.
\label{Usrmuinf2}
\end{eqnarray}
By carefully comparing Eq.~(\ref{Exsrmuinf2spin}) and Eq.~(\ref{Usrmuinf2}), one sees that, for a one-electron system ($n_{\alpha}=n$, $n_{\beta}=0$, $\tau_{\alpha}=\tau_W=|\nabla n|^2/(8n)$, $\tau_{\beta}=0$), the Hartree and exchange energies cancel out order by order in the expansion with respect to $\mu$. As the first term (in $\mu^{-2}$) of the asymptotic expansion of the exchange energy is a local functional of the density, this cancellation is exactly maintained in the LDA, i.e. there is no self-interaction error in LDA for this first term. For the next term of the expansion (in $\mu^{-4}$), the removal of the self-interaction error requires the consideration of a meta-GGA functional depending explicitly on $|\nabla n_{\sigma}|$ and $\tau_{\sigma}$.

In the KS scheme, it has been often argued that (semi)local approximate exchange functionals actually mimics near-degeneracy correlation effects (see, e.g., Ref.~\onlinecite{GriSchBae-JCP-97}). It is clear from this above discussion that, for short-range exchange functionals, (semi)local approximations become exact at large $\mu$ and therefore do not fortuitously mimic near-degeneracy correlation anymore.

  \subsection{Correlation functional for large $\mu$}

In Appendix~\ref{app:asymptoticexpansion}, we derive the first two terms in the asymptotic expansion of the short-range correlation energy for $\mu \to \infty$
\begin{equation}
\label{Ecsrmuinf2text}
\bar{E}_{c}^{\mu} = \frac{\sqrt{\pi} A_0}{\mu^2} \int n_{2,c}(\b{r},\b{r}) d\b{r} + \frac{2 \sqrt{\pi} A_1}{3 \mu^3} \int n_{2}(\b{r},\b{r}) d\b{r} +\cdots,
\end{equation}
where $n_{2,c}(\b{r},\b{r})$ and $n_{2}(\b{r},\b{r})$ are the correlation and total pair density, the $A_0$ coefficient is given after Eq.~(\ref{Exsrmuinf2spin}) and in addition $A_{1,\erf}=1$ and $A_{1,\erfgau}=28$. 

This result show that for large $\mu$ the short-range correlation energy becomes an exact local functional of the on-top pair density $n_{2}(\b{r},\b{r})$ (and of the density via $n_{2,c}(\b{r},\b{r})=n_{2}(\b{r},\b{r})-n(\b{r})n(\b{r})/2$ for closed-shell systems). This emphasizes the importance of this quantity for density functional approximations. It is in the same line of thought of a number of previous studies~\cite{MosSan-PRA-91,BecSavSto-TCA-95,PerSavBur-PRA-95} stressing the need of including explicitly the on-top pair density in approximate density functionals.

\section{Performance of the LDA for short-range exchange and correlation energies}
\label{sec:discussion}

\begin{figure}
\includegraphics[scale=0.75]{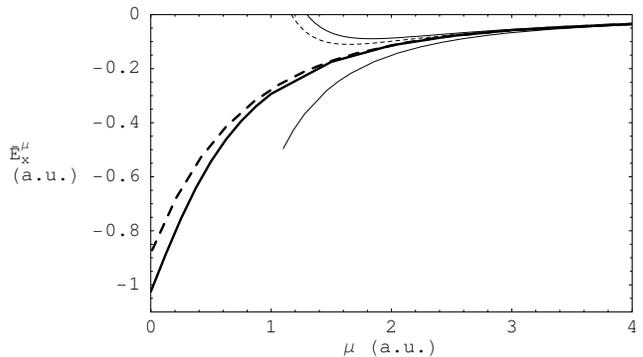}
\caption{Accurate short-range exchange energy of He (thick solid curve) along the erf adiabatic connection, local density approximation (thick long-dashed curve), exact asymptotic expansion for $\mu \to \infty$ (Eq.~\ref{Exsrmuinf2}) with the first term (lower solid curve) and with the first two terms (upper solid curve), and asymptotic expansion in LDA (Eq.~\ref{Exsrmuinf2lda}) with the first two terms (short-dashed curve), the first term in LDA being exact.
}
\label{fig:ex-he-erf-lda}
\end{figure}

\begin{figure}
\includegraphics[scale=0.75]{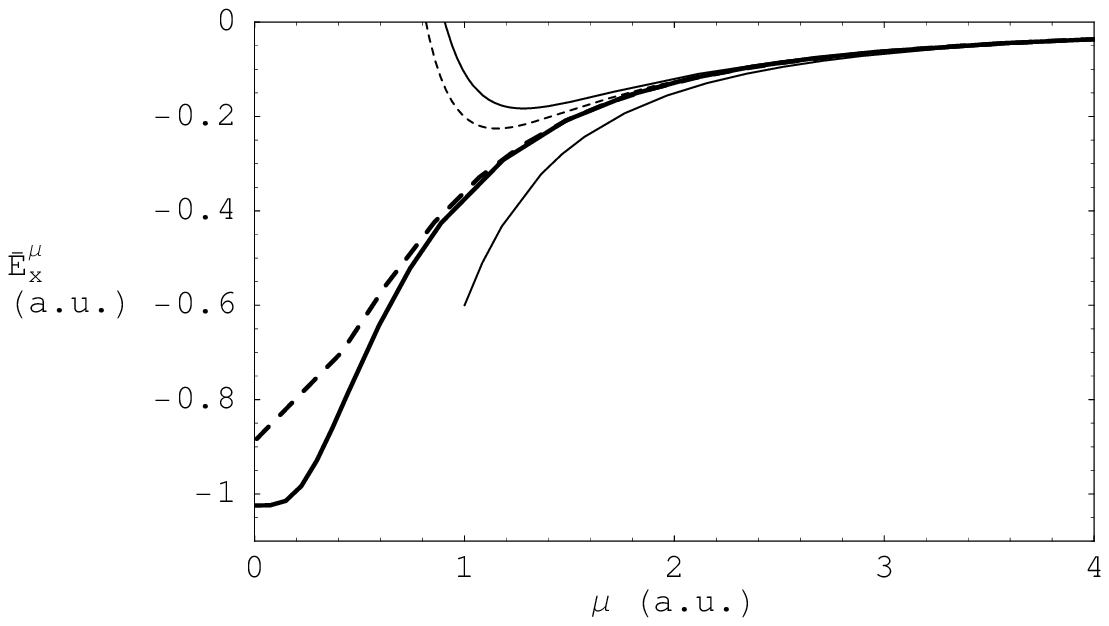}
\caption{Accurate short-range exchange energy of He (thick solid curve) along the erfgau adiabatic connection, local density approximation (thick long-dashed curve), exact asymptotic expansion for $\mu \to \infty$ (Eq.~\ref{Exsrmuinf2}) with the first term (lower solid curve) and with the first two terms (upper solid curve), and asymptotic expansion in LDA (Eq.~\ref{Exsrmuinf2lda}) with the first two terms (short-dashed curve), the first term in LDA being exact.
}
\label{fig:ex-he-erfgau-lda}
\end{figure}

The asymptotic expansions for $\mu \to \infty$ enables to analyze the local density approximation to the short-range exchange-correlation functional. We restrict the discussion to spin-unpolarized systems. Applying the LDA to the expansion of the exchange energy~(\ref{Exsrmuinftext}) corresponds to transferring the on-top exchange pair density and its derivatives $\tilde{n}_{2,x}^{(2n)}(\b{r},\b{r})$ from the uniform electron gas~\cite{GunLun-PRB-76}
\begin{equation}
\tilde{n}_{2,x}^{(2n)}(\b{r},\b{r}) \approx n(\b{r}) n(\b{r}) \left( g_{x,0}^{(2n)}(n(\b{r})) -1 \right),
\end{equation}
where $g_{x,0}^{(2n)}(n(\b{r}))$ are the on-top derivatives of the exchange-only pair-distribution function of the uniform electron gas~\cite{Raj-ACP-80}
\begin{equation}
\label{gx}
g_x(r_{12},n) = 1 - \frac{9}{2} \left( \frac{\sin(k_F r_{12}) - k_F r_{12} \cos(k_F r_{12})}{k_F^3 r_{12}^3} \right)^2,
\end{equation} 
with $k_F = (3 \pi^2 n)^{1/3}$. The expansion~(\ref{Exsrmuinf2}) of the short-range exchange energy then writes
\begin{eqnarray}
\bar{E}_{x}^{\mu,\text{LDA}} &=& -\frac{\sqrt{\pi} A_0}{2 \mu^2} \int n(\b{r})^2 d\b{r} 
\nonumber\\
&&+ \frac{3^{2/3}\pi^{11/6} A_2}{20 \mu^4} \int n(\b{r})^{8/3} d\b{r} 
+\cdots.
\label{Exsrmuinf2lda}
\end{eqnarray}
This last expansion is compared with the exact expansion~(\ref{Exsrmuinf2}) for the He atom with the erf and erfgau interactions in Figs.~\ref{fig:ex-he-erf-lda} and~\ref{fig:ex-he-erfgau-lda}, respectively. In these expressions, an accurate density of the He atom is used. The LDA for the modified interactions~\cite{Sav-INC-96,TouSavFla-IJQC-XX} evaluated with the same density is also reported, as well as an accurate calculation of the exchange energy along the adiabatic connection~\cite{PolColLeiStoWerSav-IJQC-03,TouColSav-JJJ-XXa} (see Sec.~\ref{sec:accurate}). The LDA expansion~(\ref{Exsrmuinf2lda}) is exact for the first term and very close to the exact expansion~(\ref{Exsrmuinf2}) with the first two terms. Consequently, in the domain of validity of this expansion (for $\mu \gtrsim 2$), the LDA is nearly exact. The success of the LDA for large $\mu$ is therefore due to the exactness of the exchange on-top pair-density and the good transferability of its first derivatives from the uniform electron gas to the finite system. Actually, it can be remarked that the expansion of LDA up to the $\mu^{-4}$ term performs even slightly better than the exact expansion to the same order. Similarly, the LDA is already very accurate for $\mu \gtrsim 1$ while achieving a comparable accuracy from the slowly-improving expansion~(\ref{Exsrmuinf2}) of the exact exchange energy would require a rather long expansion. 

\begin{figure}
\includegraphics[scale=0.75]{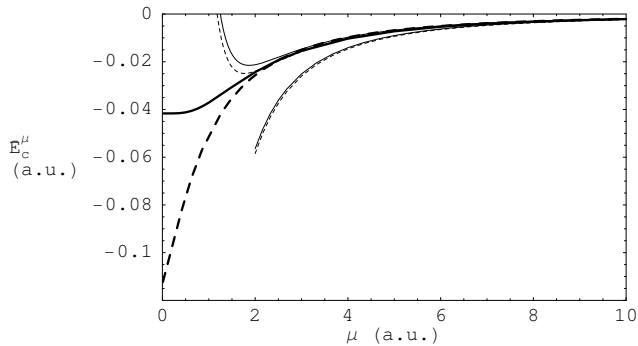}
\caption{Accurate short-range correlation energy of He (thick solid curve) along the erf adiabatic connection, local density approximation (thick long-dashed curve),
exact asymptotic expansion for $\mu \to \infty$ (Eq.~\ref{Ecsrmuinf2text}) with the first term (lower solid curve) and with the first two terms (upper solid curve), and asymptotic expansion in LDA (Eq.~\ref{Ecsrmuinf2lda}) with the first term (lower short-dashed curve) and with the first two terms (upper short-dashed curve).
}
\label{fig:ec-he-erf-lda}
\end{figure}

\begin{figure}
\includegraphics[scale=0.75]{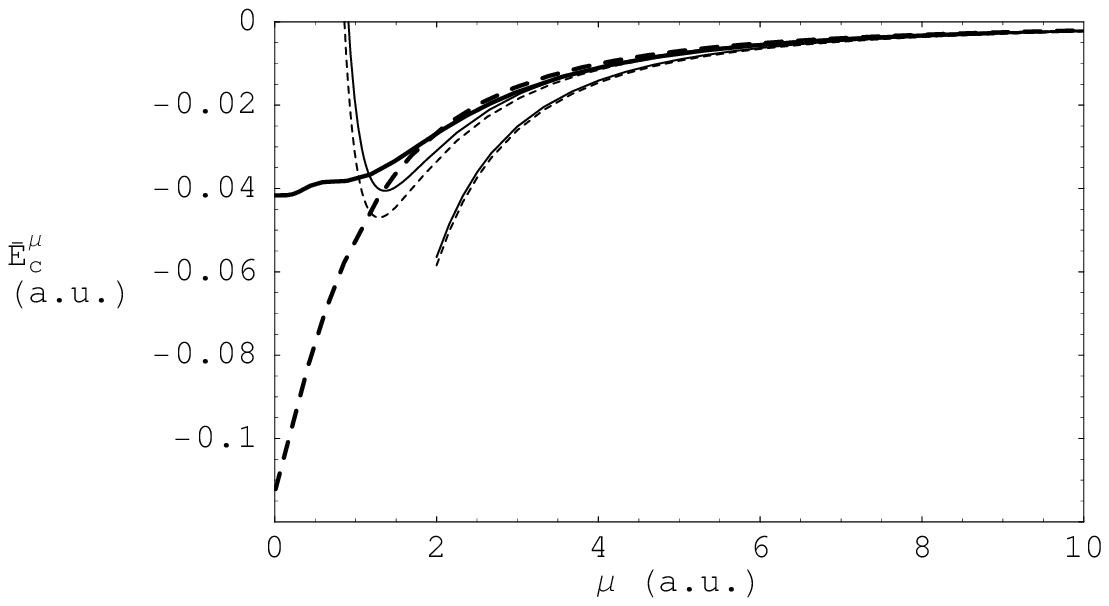}
\caption{Accurate short-range correlation energy of He (thick solid curve) along the erfgau adiabatic connection, local density approximation (thick long-dashed curve), exact asymptotic expansion for $\mu \to \infty$ (Eq.~\ref{Ecsrmuinf2text}) with the first term (lower solid curve) and with the first two terms (upper solid curve), and asymptotic expansion in LDA (Eq.~\ref{Ecsrmuinf2lda}) with the first term (lower short-dashed curve) and with the first two terms (upper short-dashed curve).
}
\label{fig:ec-he-erfgau-lda}
\end{figure}

For the correlation energy, the LDA of expansion~(\ref{Ecsrmuinf2text}) consists in transferring the on-top density from the uniform electron gas to obtain
\begin{eqnarray}
\bar{E}_{c}^{\mu,\text{LDA}} &=& \frac{\sqrt{\pi} A_0}{\mu^2} \int n(\b{r}) n(\b{r})
 \left( g_0(r_s(\b{r})) -\frac{1}{2} \right) d\b{r}
\nonumber\\
 & &+ \frac{2 A_1 \sqrt{\pi}}{3\mu^3} \int n(\b{r}) n(\b{r}) g_0(r_s(\b{r})) d\b{r} +\cdots,
\nonumber\\
\label{Ecsrmuinf2lda}
\end{eqnarray}
where $r_s=3/(4\pi n)^{1/3}$ is the local Wigner-Seitz radius and $g_0(r_s)$ is the on-top pair-distribution function of the electron gas for which Burke, Perdew and Ernzerhof have proposed an estimation~\cite{BurPerErn-JCP-98}
\begin{equation}
g_0(r_s)=D \left( (\gamma + r_s )^{3/2} +\beta \right) e^{-A \sqrt{\gamma+r_s}},
\end{equation}
with $D=32/(3\pi)$, $A=3.2581$, $\beta=163.44$ and $\gamma=4.7125$. The expansion~(\ref{Ecsrmuinf2lda}) is compared with the exact expansion~(\ref{Ecsrmuinf2text})  using an accurate calculation of $n_2(\b{r},\b{r})$ for the He atom with the erf and erfgau interactions in Figs.~\ref{fig:ec-he-erf-lda} and~\ref{fig:ec-he-erfgau-lda}, respectively. An accurate calculation of the correlation energy computed along the adiabatic connection~\cite{PolColLeiStoWerSav-IJQC-03,TouColSav-JJJ-XXa} and the LDA for modified interactions~\cite{Sav-INC-96,TouSavFla-IJQC-XX} are also reported. As for the exchange energy, the first terms of the LDA expansion~(\ref{Ecsrmuinf2lda}) nearly coincide with the exact expansion. This expansion with the first two terms gives a very accurate approximation to the exact correlation energy in the region of accuracy of the full LDA curve (from $\mu \approx 2$ to $\mu \to \infty$). From these results, it is clear that the total on-top pair density $n_2(\b{r},\b{r})$ have good transferability from the uniform electron gas to the He atom. Actually, this good (while not exact) transferability seems quite general and has already been pointed out for several atomic and molecular systems~\cite{PerSavBur-PRA-95,PerErnBurSav-IJQC-97,BurPerErn-JCP-98}. This gives an explanation for the success of the LDA in treating short-range electron-electron interactions (large $\mu$). On the contrary, toward the Kohn-Sham end ($\mu=0$) of the adiabatic connection, the LDA transfers spurious long-range correlations from the uniform electron gas and therefore poorly extrapolates the exact correlation energy of the finite system which does not contain these long-range correlation effects.

\section{Interpolations for the short-range exchange and correlation functionals}
\label{sec:approx}

In the previous section, we have shown that the LDA treats successfully short-range interactions corresponding to large interaction parameters $\mu$ but is inaccurate toward the KS end of the adiabatic connection, i.e. for small $\mu$. But for $\mu=0$ a lot of better estimates of the exchange and correlation energies are available with density functional approximations of the KS scheme which go beyond the LDA such as gradient-corrected functionals. A simple idea for improving the short-range exchange-correlation energy functional along the adiabatic connection is therefore to interpolate between an available density functional approximation (DFA) for $\mu=0$ and the $\mu$-dependent LDA for $\mu \to \infty$. In the spirit of the usual DFT approximations, this interpolation will be done locally, i.e. for the short-range exchange-correlation energy density $\bar{\varepsilon}_{xc}^{\mu}(\b{r})$ related to the global functional $\bar{E}_{xc}^{\mu}[n]$ via
\begin{equation}
\bar{E}_{xc}^{\mu}[n] = \int d\b{r} \, n(\b{r}) \, \bar{\varepsilon}_{xc}^{\mu}(\b{r}).
\end{equation}

\subsection{Rational interpolations}
\label{sec:inter}

A simple possibility is to interpolate $\bar{\varepsilon}_{xc}^{\mu}(\b{r})$ along the adiabatic connection using an estimate at $\mu=0$ and the expansions for large $\mu$ (and eventually for small $\mu$) presented in Sec.~\ref{sec:srxc}. For example, consider the rational approximant for the short-range exchange energy density of the erf interaction
\begin{equation}
\label{epsxerfinter}
\bar{\varepsilon}_{x,\erf}^{\mu} \approx \frac{\varepsilon_{x}^{\text{DFA}}}{1+b_1 \mu + b_2 \mu^2},
\end{equation}
where $b_1 = -1/(\varepsilon_{x}^{\text{DFA}}\sqrt{\pi})$ and $b_2 = -4 \varepsilon_x^{\text{DFA}}/(\pi n)$ are chosen to satisfy the expansion for small $\mu$ (Eq.~\ref{Exsrmu0erf}) and the expansion for large $\mu$ (Eq.~\ref{Exsrmuinf2}) to leading order. In Eq.~(\ref{epsxerfinter}), $\varepsilon_{x}^{\text{DFA}}$ is the exchange energy density at $\mu=0$ estimated by one of the usual density functional approximations of the KS scheme.

For the erfgau interaction, the short-range exchange energy density can also be interpolated between $\varepsilon_{x}^{\text{DFA}}$ at $\mu=0$ (no linear term in $\mu$, Eq.~\ref{Exsrmu0erfgau}) and the expansion for large $\mu$ (Eq.~\ref{Exsrmuinf2}) according to
\begin{equation}
\label{epsxerfgauinter}
\bar{\varepsilon}_{x,\erfgau}^{\mu} \approx \frac{\varepsilon_{x}^{\text{DFA}}}{1 + c_2 \mu^2},
\end{equation}
with $c_2 = -4 \varepsilon_x^{\text{DFA}}/((1+6\sqrt{3})\pi n)$. 

In the same spirit, the short-range correlation density along the adiabatic connection can be approximated by
\begin{equation}
\label{epscinter}
\bar{\varepsilon}_{c}^{\mu} \approx \frac{\varepsilon_{c}^{\text{DFA}}}{1+d_1 \mu + d_2 \mu^2},
\end{equation}
with $d_{1,\erf} = -2 d_{2,\erf}^2 \sqrt{\pi} n g_0(r_s) /(3 \varepsilon_{c})$, $d_{2,\erf} = 2 \varepsilon_{c} /(\pi n (g_0(r_s)-1/2))$ for the erf interaction, and $d_{1,\erfgau} = -56 d_{2,\erfgau}^2 \sqrt{\pi} n g_0(r_s) /(3 \varepsilon_{c})$, $d_{2,\erfgau} = d_{2,\erf}/(1+6\sqrt{3})$ for the erfgau interaction. Again, $\varepsilon_{c}^{\text{DFA}}$ is the correlation energy density at $\mu=0$ given by an available density functional approximation of the KS scheme. The rational approximant~(\ref{epscinter}) therefore interpolates between $\varepsilon_{c}^{\text{DFA}}$ at $\mu=0$ and the expansion of the LDA for large $\mu$ with the first two terms (Eq.~\ref{Ecsrmuinf2lda}). 

\subsection{Weighted interpolations}
\label{sec:inter2}

The interpolation formulas~(\ref{epsxerfinter}) and~(\ref{epsxerfgauinter}) only make use of the first term of the asymptotic expansion of the LDA exchange energy for $\mu \to \infty$. However, it has been realized in Sec.~\ref{sec:discussion} that the LDA works well on a larger domain of $\mu$ than the first terms of its asymptotic expansion. In order to take better advantage of the LDA, one can modify it only in the region of small $\mu$ by using information from better estimates of the KS exchange energy at $\mu=0$. We therefore interpolate locally the short-range exchange density along the adiabatic connection by
\begin{equation}
\label{epsxinter2}
\bar{\varepsilon}_{x}^{\mu} \approx (\varepsilon_{x}^{\text{DFA}} - \bar{\varepsilon}_{x}^{\mu=0,\text{unif}}) w(\mu) + \bar{\varepsilon}_{x}^{\mu,\text{unif}},
\end{equation}
where $w(\mu)$ is a weight function acting at small $\mu$ only. More precisely, $w(\mu)$ must be a positive function satisfying $w(\mu=0)=1$ so that $\bar{\varepsilon}_{x}^{\mu=0} = \varepsilon_{x}^{\text{DFA}}$, having significant values in the region where the LDA fails and with a fast decay for $\mu \to \infty$ so as to recover the correct behavior of $\bar{\varepsilon}_{x}^{\mu,\text{unif}}$ for large $\mu$. For the erf interaction, we found that the local value of $\mu$ delimiting the domain where the LDA is inaccurate is well estimated by $1/r_s(\b{r})$ where $r_s(\b{r})$ is the local Wigner-Seitz radius. We therefore take $w_{\erf}(\mu)=\erfc(\mu r_s)$ where the complementary error function $\erfc$ ensures that the weight function has significant values only for $\mu < 1/r_s$. Eq.~(\ref{epsxinter2}) thus interpolates between $\varepsilon_{x}^{\text{DFA}}$ at $\mu=0$ and the LDA at large $\mu$.

For the erfgau interaction, we can used the same weight function except that now the local value of $\mu$ delimiting the domain where the LDA must be corrected is estimated by $\mu=c/r_s$ where $c = \left( 1+6\sqrt{3}\right)^{1/2} \approx 3.375$ is the scale factor between the erf and erfgau interactions discussed in Sec.~\ref{sec:lrsrseparation}. We therefore take as weight function $w_{\erfgau}(\mu)=\erfc(\mu r_s/c)$.

Naturally, this interpolation can also be applied to the short-range correlation energy
\begin{equation}
\label{epscinter2}
\bar{\varepsilon}_{c}^{\mu} \approx (\varepsilon_{c}^{\text{DFA}} - \bar{\varepsilon}_{c}^{\mu=0,\text{unif}}) w(\mu) + \bar{\varepsilon}_{c}^{\mu,\text{unif}},
\end{equation}
with the same weight function $w(\mu)$.

\subsection{Other interpolations}
\label{sec:interHirao}

Finally, we mention the approximation for the short-range exchange energy proposed by Iikura, Tsuneda, Yanai and Hirao~\cite{IikTsuYanHir-JCP-01} based on a modification of the short-range LDA exchange functional~\cite{Sav-INC-96,TouSavFla-IJQC-XX}. The spin-unpolarized version of their approximation for the erf interaction is
\begin{eqnarray}
\bar{\varepsilon}_{x,\erf}^{\mu} &\approx& \varepsilon_{x}^{\text{DFA}} \Biggl[ 1 - \frac{8}{3} A \biggl( \sqrt{\pi} \erf \left( \frac{1}{2 A} \right)
\nonumber\\
&& + (2 A - 4 A^3) e^{-1/(4 A^2)} - 3 A + 4 A^3 \biggl) \Biggl],
\label{epsxerfinterHirao}
\end{eqnarray}
where $A= \mu/(2 k)$, $k=\sqrt{\varepsilon_{x}^{\text{LDA}}/\varepsilon_{x}^{\text{DFA}}} k_F$ and $k_F=(3\pi^2 n)^{1/3}$. The approximation reduces to $\varepsilon_{x}^{\text{DFA}}$ at $\mu=0$ and has an asymptotic expansion for $\mu \to \infty$ incorporating the correct leading term (cf. Eq.~\ref{Exsrmuinf2}). Therefore, Eq.~(\ref{epsxerfinterHirao}) provides an interpolation between a density functional approximation at $\mu=0$, $\varepsilon_{x}^{\text{DFA}}$, and the correct limit as $\mu \to \infty$. The same approximation can also be derived for the erfgau interaction using the LDA exchange functional associated to this interaction~\cite{TouSavFla-IJQC-XX}; it reads 

\begin{eqnarray}
\bar{\varepsilon}_{x,\erfgau}^{\mu} &\approx& \varepsilon_{x}^{\text{DFA}} \Biggl[ 1 - \frac{8}{3} A \biggl( \sqrt{\pi} \erf \left( \frac{1}{2 A} \right)
\nonumber\\
&& + (2 A - 4 A^3) e^{-1/(4 A^2)} - 3 A + 4 A^3 \biggl) 
\nonumber\\
&& + \frac{8}{3} A \biggl( \sqrt{\pi} \erf \left( \frac{1}{2 B} \right)
\nonumber\\
&& + (2 B - 16 B^3) e^{-1/(4 B^2)} - 6 B + 16 B^3 \biggl) \Biggl],
\nonumber\\
\label{epsxerfgauinterHirao}
\end{eqnarray}
where $B=A/\sqrt{3}$. 

Notice that in the interpolations of Sec.~\ref{sec:inter},~\ref{sec:inter2} and~\ref{sec:interHirao} the exchange and correlation energy densities at $\mu=0$, $\varepsilon_{x}^{\text{DFA}}$ and $\varepsilon_{c}^{\text{DFA}}$, can be estimated by any of the available exchange-correlation functionals of the Kohn-Sham scheme. In this sense, formulas~(\ref{epsxerfinter}) to~(\ref{epsxerfgauinterHirao}) provide extensions of these exchange-correlation functionals over the erf and erfgau adiabatic connections.

\subsection{Results}

We now test the interpolations formulas~(\ref{epsxerfinter}) to~(\ref{epsxerfgauinterHirao}) on a few atomic systems. In all the results presented here, we use the PBE functional~\cite{PerBurErn-PRL-96} as the density functional approximation at $\mu=0$ both for exchange and correlation: $\varepsilon_{x}^{\text{DFA}}=\varepsilon_{x}^{\text{PBE}}$ and $\varepsilon_{c}^{\text{DFA}}=\varepsilon_{c}^{\text{PBE}}$.

\begin{figure}
\includegraphics[scale=0.75]{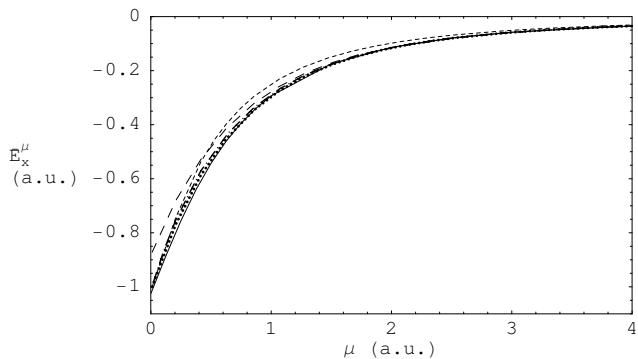}
\caption{Accurate short-range exchange energy of He (solid curve) along the erf adiabatic connection, local density approximation (long-dashed curve) and three interpolations between the PBE value at $\mu=0$ and the LDA at $\mu \to \infty$ using a rational approximant (Eq.~\ref{epsxerfinter}, short-dashed curve), a weighted approximant (Eq.~\ref{epsxinter2}, dotted curve) and the approximation of Iikura \textit{et al.} (Eq.~\ref{epsxerfinterHirao}, dashed-dotted curve). The last two curves are nearly superimposed.
}
\label{fig:ex-he-erf-inter}
\end{figure}

\begin{figure}
\includegraphics[scale=0.75]{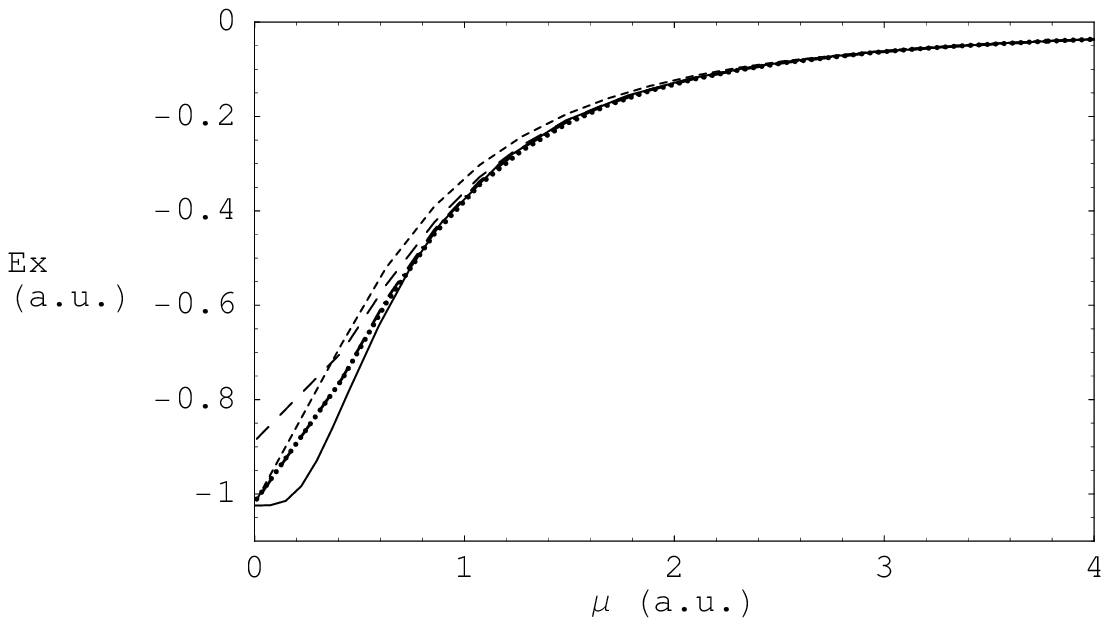}
\caption{Accurate short-range exchange energy of He (solid curve) along the erfgau adiabatic connection, local density approximation (long-dashed curve) and three interpolations between the PBE value at $\mu=0$ and the LDA at $\mu \to \infty$ using a rational approximant (Eq.~\ref{epsxerfgauinter}, short-dashed curve), a weighted approximant (Eq.~\ref{epsxinter2}, dotted curve) and the approximation of Iikura \textit{et al.} (Eq.~\ref{epsxerfgauinterHirao}, dashed-dotted curve). The last two curves are nearly superimposed.
}
\label{fig:ex-he-erfgau-inter}
\end{figure}

Figs.~\ref{fig:ex-he-erf-inter} and~\ref{fig:ex-he-erfgau-inter} represent the short-range exchange energy of the He atom along the erf and erfgau adiabatic connections, respectively. An accurate calculation is compared to the LDA and to the three interpolations of Sec.~\ref{sec:inter},~\ref{sec:inter2} and~\ref{sec:interHirao}. One sees that the rational approximants give an overall reasonable estimate of the exchange energy along both adiabatic connections but are actually less accurate that the LDA in the region of intermediate $\mu$. As already noticed, this reflects the fact that the LDA works well on a larger range of $\mu$ that the first terms of its expansion for $\mu \to \infty$. The weighted interpolations make better use of the LDA and constitutes an improvement over it for all $\mu$'s. Finally, the approximation of Iikura \textit{et al.} is nearly identical to the weighted interpolation for both the erf and erfgau interactions.

\begin{figure}
\includegraphics[scale=0.75]{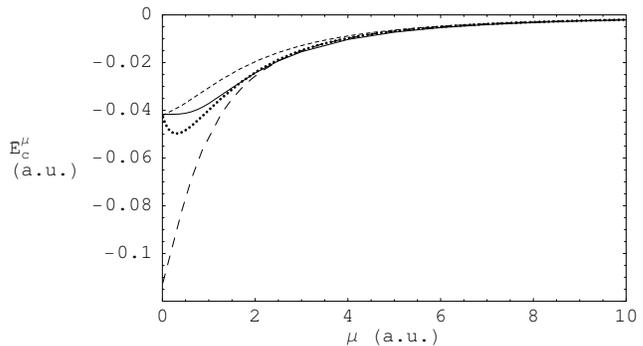}
\caption{Accurate short-range correlation energy of He (solid curve) along the erf adiabatic connection, local density approximation (long-dashed curve) and two interpolations between the PBE value at $\mu=0$ and the LDA at $\mu \to \infty$ using a rational approximant (Eq.~\ref{epscinter}, short-dashed curve) and a weighted approximant (Eq.~\ref{epscinter2}, dotted curve).
}
\label{fig:ec-he-erf-inter}
\end{figure}

\begin{figure}
\includegraphics[scale=0.75]{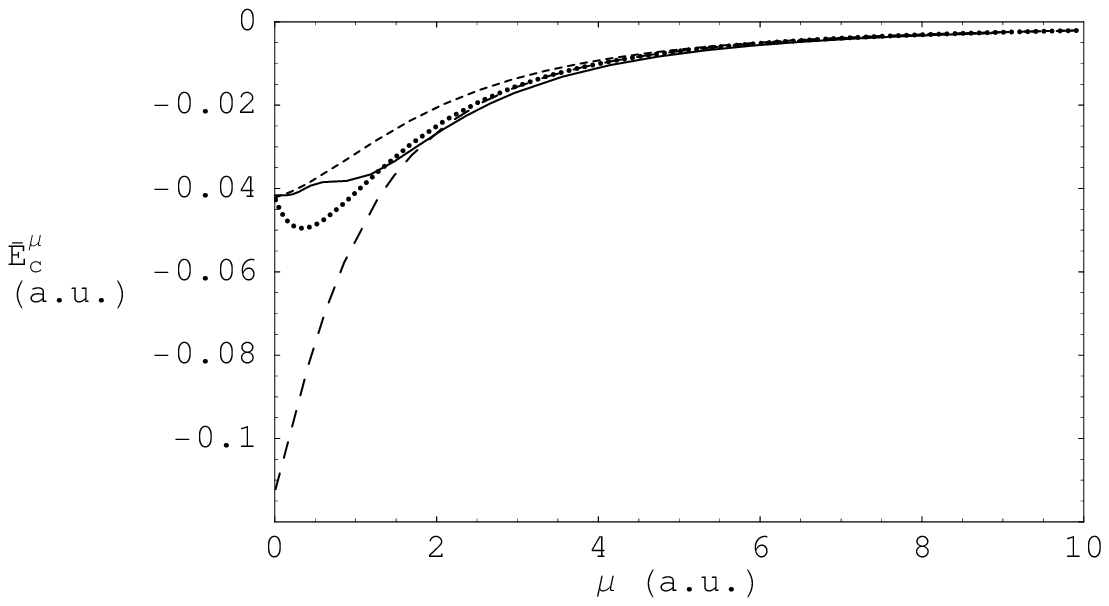}
\caption{Accurate short-range correlation energy of He (solid curve) along the erfgau adiabatic connection, local density approximation (long-dashed curve) and two interpolations between the PBE value at $\mu=0$ and the LDA at $\mu \to \infty$ using a rational approximant (Eq.~\ref{epscinter}, short-dashed curve) and a weighted approximant (Eq.~\ref{epscinter2}, dotted curve).
}
\label{fig:ec-he-erfgau-inter}
\end{figure}

Similar curves for the short-range correlation energy of the He atom with the erf and erfgau interactions are shown in Figs.~\ref{fig:ec-he-erf-inter} and~\ref{fig:ec-he-erfgau-inter}. As for the exchange energy, the rational approximants constitute an overall correction to the LDA but not in a systematic way since the LDA still performs better for intermediate $\mu$. On the contrary, the weighted interpolations always improve the LDA. As one can expect, considering exchange and correlation together further improves the results at small $\mu$.

\begin{figure}
\includegraphics[scale=0.75]{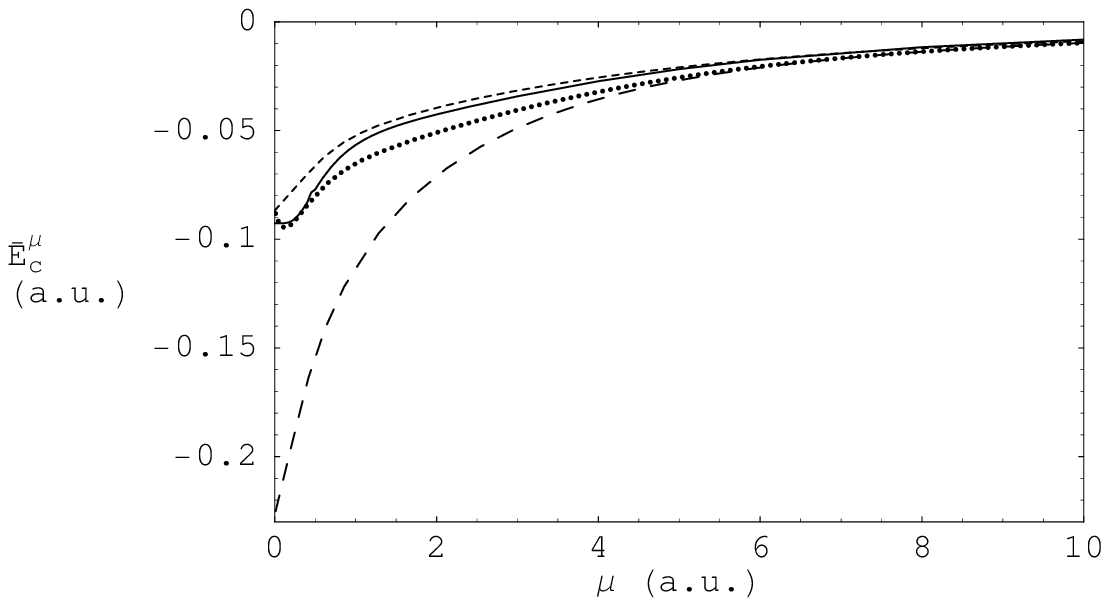}
\caption{Accurate short-range correlation energy of Be (solid curve) along the erf adiabatic connection, local density approximation (long-dashed curve) and two interpolations between the PBE value at $\mu=0$ and the LDA at $\mu \to \infty$ using a rational approximant (Eq.~\ref{epscinter}, short-dashed curve) and a weighted approximant (Eq.~\ref{epscinter2}, dotted curve).
}
\label{fig:ec-be-erf-inter}
\end{figure}

\begin{figure}
\includegraphics[scale=0.75]{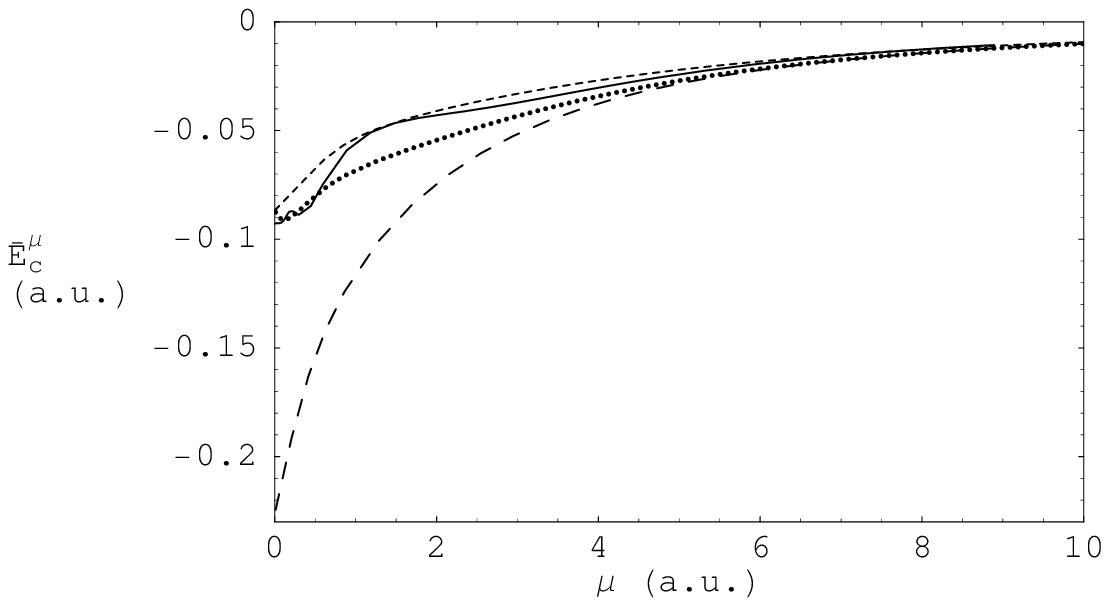}
\caption{Accurate short-range correlation energy of Be (solid curve) along the erfgau adiabatic connection, local density approximation (long-dashed curve) and two interpolations between the PBE value at $\mu=0$ and the LDA at $\mu \to \infty$ using a rational approximant (Eq.~\ref{epscinter}, short-dashed curve) and a weighted approximant (Eq.~\ref{epscinter2}, dotted curve).
}
\label{fig:ec-be-erfgau-inter}
\end{figure}

We now discuss the Be atom. The short-range exchange energy along the adiabatic connections looks very similar to that of the He atom and will not be shown. The short-range correlation energy for which the LDA has more difficulties than for the He case is represented in  Figs.~\ref{fig:ec-be-erf-inter} and~\ref{fig:ec-be-erfgau-inter} for the erf and erfgau interactions. One can see that both the rational approximant and the weighted interpolation improve the LDA along the whole adiabatic connection.

\begin{figure}
\includegraphics[scale=0.75]{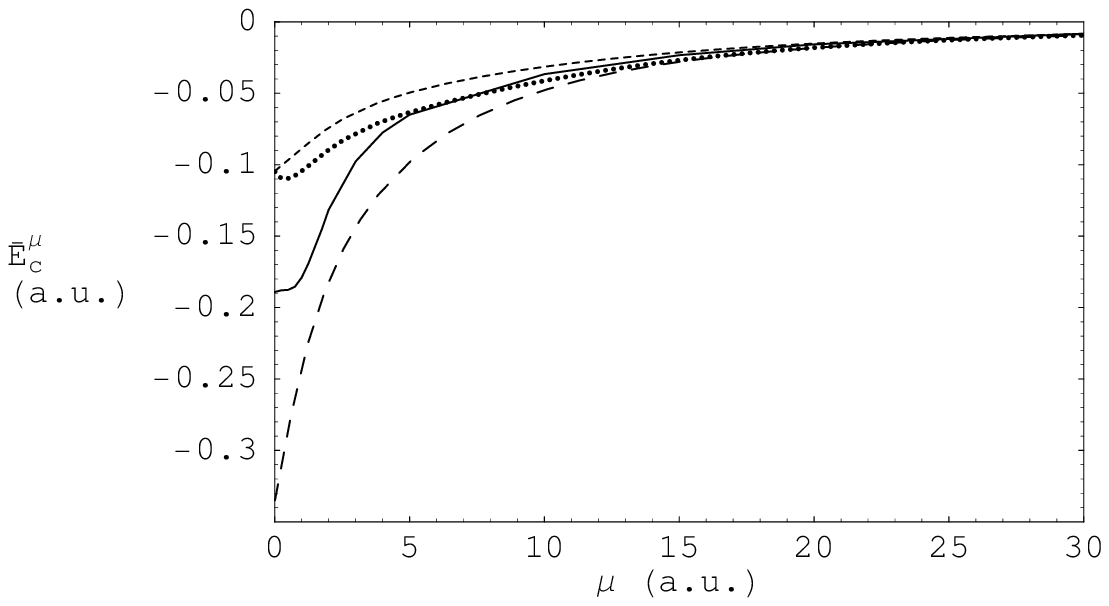}
\caption{Accurate short-range correlation energy of Ne$^{6+}$ (solid curve) along the erf adiabatic connection, local density approximation (long-dashed curve) and two interpolations between the PBE value at $\mu=0$ and the LDA at $\mu \to \infty$ using a rational approximant (Eq.~\ref{epscinter}, short-dashed curve) and a weighted approximant (Eq.~\ref{epscinter2}, dotted curve).
}
\label{fig:ec-ne6p-erf-inter}
\end{figure}

\begin{figure}
\includegraphics[scale=0.75]{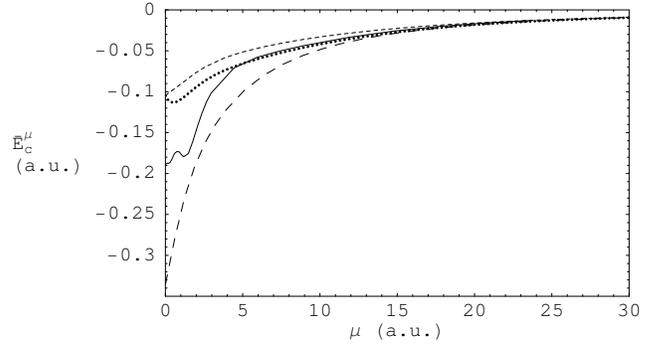}
\caption{Accurate short-range correlation energy of Ne$^{6+}$ (solid curve) along the erfgau adiabatic connection, local density approximation (long-dashed curve) and two interpolations between the PBE value at $\mu=0$ and the LDA at $\mu \to \infty$ using a rational approximant (Eq.~\ref{epscinter}, short-dashed curve) and a weighted approximant (Eq.~\ref{epscinter2}, dotted curve).
}
\label{fig:ec-ne6p-erfgau-inter}
\end{figure}

The He and Be atoms are simple cases where a gradient-corrected functional like PBE give a very accurate correlation energy. On the contrary, the Ne$^{6+}$ atom constitutes a much more difficult system for (semi)local functionals because of the presence of strong near-degeneracy correlation effects due to the proximity of the $2s$ and $2p$ levels. The short-range correlation energy of this system is reported in Figs.~\ref{fig:ec-ne6p-erf-inter} and~\ref{fig:ec-ne6p-erfgau-inter} for the erf and erfgau interactions. The PBE functional of the KS scheme ($\mu=0$) strongly underestimates the correlation energy. One sees that with the rational approximants or the weighted interpolations, the error is rapidly decreased when $\mu$ is increased, i.e. when long-range interactions are removed from the functional.

\section{Conclusion}
\label{sec:conclusion}

In a quantum electronic system, the long-range/short-range separation of the Coulomb interaction enables to rigorously decompose the total energy into long-range and short-range components which can be calculated by different methods. In particular, a density functional approximation can be used for the short-range part of the energy, while the long-range contribution can be treated by traditional wave function methods. In this work, we have considered two possible long-range/short-range separation of the Coulomb interaction: the erf and erfgau modified interactions. The erfgau interaction achieves a better separation than the erf interaction. We have shown that the use of these modified interactions facilitates wave function calculations. We have also studied the short-range part of the exchange-correlation functional with respect to the range of the associated interaction, and shown that, in the limit of a very short-range interaction, the exchange contribution to this functional can be expressed as a local functional of the density and is therefore exact in the LDA. In the same limit, the correlation contribution can be expressed as a local functional of the on-top pair density which is generally accurate in the LDA. However, when the interaction becomes more and more long-ranged, the LDA is inaccurate. It is nevertheless possible to improve the LDA description of the short-range exchange-correlation functional by making use of the available gradient-corrected functional of the Kohn-Sham scheme in the limit of the full Coulomb interaction. We have indeed proposed two kind of interpolations for the short-range exchange-correlation functional along the erf or erfgau adiabatic connection which improve the LDA. In order to extend the interaction range well treated by functional approximations in a more systematic way, we are currently investigating gradient corrections for the short-range exchange-correlation functional.

\begin{acknowledgments}
We are grateful to P. Gori-Giorgi (Universit\'e Paris VI, France) for stimulating discussions and in particular for suggesting us Eq.~(\ref{psimuerf}).
\end{acknowledgments}

\appendix

\begin{widetext}
\section{Modified two-electron integrals}
\label{app:integrals}
In this Appendix, we give details of the evaluation of the two-electron integrals over Gaussian basis sets for the modified erf and erfgau interactions. The modified integrals have been implemented into the Seward program~\cite{LinRyuLiu-JCP-91} available in the Molpro package~\cite{Molproshort-PROG-02} and using the Rys quadrature scheme~\cite{DupRysKin-JCP-76}. For other implementations of two-electron integrals with modified interactions, see Refs.~\onlinecite{GilAda-CPL-96,AdaDomGil-JCC-99,PreNolFilFahGre-JCP-01}.

The general form of four-center two-electron integrals is
\begin{equation}
\label{abcd}
[\b{a} \b{b} | v_{ee}(r_{12}) | \b{c} \b{d}] = \iint d\b{r}_1 d\b{r}_2 \phi(\b{r}_1,\zeta_a,\b{a},\b{A}) \phi(\b{r}_1,\zeta_b,\b{b},\b{B}) v_{ee}(r_{12}) \phi(\b{r}_2,\zeta_c,\b{c},\b{C}) \phi(\b{r}_2,\zeta_d,\b{d},\b{D}),
\end{equation}
where $v_{ee}(r_{12})$ is the electron-electron interaction and $\phi$ is  is an unnormalized primitive Cartesian Gaussian basis function
\begin{equation}
\phi(\b{r},\zeta,\b{n},\b{R}) = (x-R_x)^{n_x} (y-R_y)^{n_y} (z-R_z)^{n_z} e^{-\zeta (\b{r}-\b{R})^2}.
\end{equation}
Actually, it is only required to explicitly compute the integrals of the type
\begin{equation}
\begin{split}
[\b{e} \b{0} | v_{ee}(r_{12}) | \b{f} \b{0}] =
\kappa_{AB} \kappa_{CD} \iint d\b{r}_1 d\b{r}_2 (x_1 - A_x)^{e_x} (y_1 - A_y)^{e_y} (z_1 - A_z)^{e_z} e^{-\zeta (\b{r}_1 - \b{P})^2} v_{ee}(r_{12}) \\
(x_2 - C_x)^{f_x} (y_2 - C_y)^{f_y} (z_2 - C_z)^{f_z} e^{-\eta (\b{r}_2 - \b{Q})^2},
\end{split}
\end{equation}
where $\zeta = \zeta_a +\zeta_b$, $\eta = \zeta_c +\zeta_d$, $\b{P}=(\zeta_a \b{A} + \zeta_b \b{B})/(\zeta_a +\zeta_b)$, $\b{Q}=(\zeta_c \b{C} + \zeta_d \b{D})/(\zeta_c +\zeta_d)$, $\kappa_{AB}=\exp\{- [\zeta_a \zeta_b / (\zeta_a +\zeta_b)](\b{A}-\b{B})^2 \}$ and $\kappa_{CD}=\exp\{- [\zeta_c \zeta_d / (\zeta_c +\zeta_d)](\b{C}-\b{B})^2 \}$. The general integrals $[\b{a} \b{b} | v_{ee}(r_{12}) | \b{c} \b{d}]$ are then obtained by applying the so-called transfer relation~\cite{LinRyuLiu-JCP-91} to the integrals $[\b{e} \b{0} | v_{ee}(r_{12}) | \b{f} \b{0}]$ with $\b{e}=\max(\b{a},\b{b})...\b{a}+\b{b}$ and $\b{f}=\max(\b{c},\b{d})...\b{c}+\b{d}$.

For the Coulomb interaction $v_{ee}(r_{12})=1/r_{12}$, these integrals can be written as
\begin{equation}
\label{intPt}
[\b{e} \b{0} | 1/r_{12} | \b{f} \b{0}] = \int_{0}^{1} dt P_n(t) e^{-t^2 \rho (\b{P}-\b{Q})^2},
\end{equation}
with $\rho=\zeta \eta/(\zeta+\eta)$ and $P_n(t)$ are the Rys polynomials of order $n=e_x + e_y + e_z + f_x + f_y + f_z$ in $t^2$
\begin{equation}
P_n(t)=2 \left(\frac{\rho}{\pi}\right)^{1/2} \kappa_{AB} \kappa_{CD} \left(\frac{\rho}{\zeta}\right)^{3/2}  \left(\frac{\rho}{\eta}\right)^{3/2} I_x'(e_x,f_x,t) I_y'(e_y,f_y,t) I_z'(e_z,f_z,t)
\end{equation}
where $I_{\lambda}'$ ($\lambda=x,y,z$) are 2D integrals obeying the recurrence relation
\begin{eqnarray}
I_{\lambda}'(e_{\lambda} +1, f_{\lambda}) &=& \left( (P_{\lambda} - A_{\lambda}) + \frac{\rho t^2}{\zeta}(Q_{\lambda} - P_{\lambda}) \right) I_{\lambda}'(e_{\lambda}, f_{\lambda}) 
\nonumber\\
&&+  \frac{e_{\lambda}}{2\zeta} \left( 1 - \frac{\rho t^2}{\zeta} \right) I_{\lambda}'(e_{\lambda} -1, f_{\lambda}) + \frac{f_{\lambda} \rho t^2}{2 \zeta \eta} I_{\lambda}'(e_{\lambda}, f_{\lambda} -1),
\label{Irecurrence}
\end{eqnarray}
with the starting value $I_{\lambda}'(0,0)=1$. Equation~(\ref{intPt}) can be evaluated exactly by the Rys quadrature
\begin{equation}
[\b{e} \b{0} | 1/r_{12} | \b{f} \b{0}] = 2 \left(\frac{\rho}{\pi}\right)^{1/2} \kappa_{AB} \kappa_{CD} \left(\frac{\rho}{\zeta}\right)^{3/2}  \left(\frac{\rho}{\eta}\right)^{3/2} \sum_{\alpha=1}^{n_{Rys}}  I_x'(e_x,f_x,t_{\alpha}) I_y'(e_y,f_y,t_{\alpha}) I_z'(e_z,f_z,t_{\alpha}) w_{\alpha},
\label{intefcoul}
\end{equation}
where $t_{\alpha}$ and $w_{\alpha}=\exp\{ -t_{\alpha}^2 \rho (\b{P}-\b{Q})^2\}$ are the roots and the weights of Rys polynomials and $n_{Rys}>n/2$.

The modified integrals for the erf interaction $[\b{e} \b{0} | \erf(\mu r_{12})/r_{12} | \b{f} \b{0}]$ can simply be computed with the same scheme by applying the following simple modification everywhere~\cite{Sav-INC-96,LeiStoWerSav-CPL-97}: $1/\rho \to 1/\rho + 1/\mu^2$. Thus, with respect to the Coulomb case, in Eq.~(\ref{Irecurrence}) the recurrence coefficients are modified, and in Eq.~(\ref{intefcoul}) the prefactor, the weights $w_{\alpha}$ and the roots $t_{\alpha}$, depending on $\rho$, are modified.

To calculate the integrals for the erfgau interaction, additional integrals of type $[\b{e} \b{0} | C \exp( -a r_{12}^2 ) | \b{f} \b{0}]$ are needed. These integrals can be computed like an one-point Rys quadrature with root $t_a =\sqrt{a^2/(\rho+a^2)}$, weight $w_a=\exp\{ -t_{a}^2 \rho (\b{P}-\b{Q})^2\}$ and a modified prefactor depending on $t_a$
\begin{eqnarray}
[\b{e} \b{0} | C \exp( -a r_{12}^2 ) | \b{f} \b{0}] &=& C (1-t_a^2)^{3/2} \kappa_{AB} \kappa_{CD} \left(\frac{\rho}{\zeta}\right)^{3/2}  \left(\frac{\rho}{\eta}\right)^{3/2}  
\nonumber\\
&&\times I_x'(e_x,f_x,t_{a}) I_y'(e_y,f_y,t_{a}) I_z'(e_z,f_z,t_{a}) w_{a}.
\end{eqnarray}
\end{widetext}

\section{Adiabatic connection near the KS system}
\label{app:expansionsmallmu}

In this appendix, we study the erf and erfgau adiabatic connections near the KS system, i.e. for small interaction parameter $\mu$. The fictitious system along these connections is described by the Hamiltonian [cf. Eq.~(\ref{HmuPsimu})]
\begin{equation}
 \hat{H}^{\mu} = \sum_{i} \frac{-1}{2} \nabla_i^2 + \sum_{i<j} v_{ee}^{\mu}(r_{ij}) + \sum_{i} v^{\mu}(\b{r}_i),
\label{Hmu}
\end{equation}
where $v_{ee}^{\mu}$ is the long-range erf or erfgau interaction and $v^{\mu}$ is the external local potential associated to the short-range Hartree, exchange and correlation energy functionals $v^{\mu}(\b{r})=v_{ne}(\b{r}) + \delta \bar{U}^{\mu}[n]/\delta n(\b{r}) + \bar{E}_{x}^{\mu}[n]/\delta n(\b{r}) + \bar{E}_{c}^{\mu}[n]/\delta n(\b{r})$. The behavior for $\mu \to 0$ of all quantities associated to this fictitious system can be derived from the Maclaurin series of the long-range interaction $v_{ee}^{\mu}$
\begin{equation}
v_{ee}^{\mu}(r_{12}) = \frac{2}{\sqrt{\pi}} \sum_{n=0}^{\infty} \frac{(-1)^n a_n}{n!} r_{12}^{2n}\mu^{2n+1},
\label{veemclaurin}
\end{equation}
where $a_{n,\erf}=1/(2n+1)$ for the erf interaction and $a_{n,\erfgau}=1/(2n+1) - 1/3^n$ ( $=0$ for $n \leq 1$ by construction) for the erfgau interaction. 

\subsection{Short-range Hartree energy for $\mu \to 0$}
Eq.~(\ref{veemclaurin}) leads immediately to the expansion of the short-range Hartree energy
\begin{eqnarray}
\bar{U}^{\mu} &=& \frac{1}{2} \iint n(\b{r}_1)n(\b{r}_2) \bar{v}_{ee}^{\mu}(r_{12}) d\b{r}_1 d\b{r}_2
\nonumber\\
 &=& U - \frac{1}{\sqrt{\pi}} \sum_{n=0}^{\infty} \frac{(-1)^n a_n}{n!} \mu^{2n+1} \nonumber\\
 & & \times \iint n(\b{r}_1)n(\b{r}_2) r_{12}^{2n} d\b{r}_1 d\b{r}_2,
\label{Usrmu0}
\end{eqnarray}
where $U$ is the standard KS Hartree energy. To obtain Eq.~(\ref{Usrmu0}), the integral and summation signs have been interchanged. This is reasonable for a finite system where $r_{12}$ is always bounded since in this case the series~(\ref{veemclaurin}) is uniformly convergent and can thus be integrated term by term. 

\subsection{Short-range exchange energy for $\mu \to 0$}
The expansion for $\mu \to 0$ of the short-range exchange energy of Eq.~(\ref{Ex}) is
\begin{eqnarray}
\bar{E}_{x}^{\mu} &=& \frac{1}{2} \iint n_{2,x}(\b{r}_1,\b{r}_{2}) \bar{v}_{ee}^{\mu}(r_{12}) d\b{r}_1 d\b{r}_{2}
\nonumber\\
 &=& E_{x} - \frac{1}{\sqrt{\pi}} \sum_{n=0}^{\infty} \frac{(-1)^n a_n}{n!} \mu^{2n+1}
\nonumber\\
 & & \times \iint n_{2,x}(\b{r}_1,\b{r}_2) r_{12}^{2n} d\b{r}_1 d\b{r}_2,
\label{Exsrmu0}
\end{eqnarray}
where $E_{x}$ is the standard KS exchange energy and $n_{2,x}(\b{r}_1,\b{r}_2)$ is the exchange contribution to the pair density. 

\subsection{Short-range correlation energy for $\mu \to 0$}

The short-range correlation energy of Eq.~(\ref{Ec}) can be written with the correlation contribution to the pair density $n_{2,c}^{\mu}(\b{r}_1,\b{r}_{2})$ with interaction $v_{ee}^{\mu}$ as
\begin{equation}
\bar{E}_{c}^{\mu}= E_c - \frac{1}{2} \iint n_{2,c}^{\mu}(\b{r}_1,\b{r}_{2}) v_{ee}^{\mu}(r_{12}) d\b{r}_1 d\b{r}_{2},
\end{equation}
where $E_c$ is the usual correlation energy of the KS scheme. The expansion of $\bar{E}_{c}^{\mu}$ for small $\mu$ can be obtained by first studying the derivative of $\bar{E}_{c}^{\mu}$ with respect to $\mu$ which by the Hellmann-Feynman theorem writes
\begin{equation}
\label{dEcdmu}
\frac{\partial \bar{E}_{c}^{\mu}}{\partial \mu} = -\frac{1}{2} \iint n_{2,c}^{\mu}(\b{r}_1,\b{r}_2) \frac{\partial v_{ee}^{\mu}(r_{12})}{\partial \mu} d\b{r}_1 d\b{r}_2,
\end{equation}
and assuming for $n_{2,c}^{\mu}(\b{r}_1,\b{r}_2)$ the following expansion around $\mu=0$
\begin{equation}
n_{2,c}^{\mu}(\b{r}_1,\b{r}_2) = \sum_{k=1}^{\infty} \frac{1}{k!} \left( \frac{\partial^k n_{2,c}^{\mu}(\b{r}_1,\b{r}_2)}{\partial \mu^k} \right)_{\mu=0}\mu^k,
\label{Pcseriesmu}
\end{equation}
since for the KS system $n_{2,c}^{\mu=0}(\b{r}_1,\b{r}_2)=0$. Therefore, inserting equation (\ref{Pcseriesmu}) and the derivative of Eq.~(\ref{veemclaurin}) into~(\ref{dEcdmu}), and assuming the commutativity of summation and integration, leads to
\begin{eqnarray}
\frac{\partial \bar{E}_{c}^{\mu}}{\partial \mu} &=& - \frac{1}{\sqrt{\pi}} \sum_{n=0}^{\infty} \sum_{k=1}^{\infty} \frac{(-1)^n (2n+1) a_n}{n! k!} \mu^{2n+k}
\nonumber\\
 & &\times \iint \left( \frac{\partial^k n_{2,c}^{\mu}(\b{r}_1,\b{r}_2)}{\partial \mu^k} \right)_{\mu=0} r_{12}^{2n} d\b{r}_1 d\b{r}_2,
\nonumber\\
\end{eqnarray}
where the term $n=0$ can be dropped since $n_{2,c}^{\mu}(\b{r}_1,\b{r}_2)$ integrates to zero. After integration we obtain the expansion of the short-range correlation energy
\begin{eqnarray}
\bar{E}_{c}^{\mu} &=& E_c - \frac{1}{\sqrt{\pi}} \sum_{n=1}^{\infty} \sum_{k=1}^{\infty} \frac{(-1)^n (2n+1) a_n}{n! k! (2n+k+1)} \mu^{2n+k+1} 
\nonumber\\
 & & \times \iint \left( \frac{\partial^k n_{2,c}^{\mu}(\b{r}_1,\b{r}_2)}{\partial \mu^k} \right)_{\mu=0} r_{12}^{2n} d\b{r}_1 d\b{r}_2.
\label{Ecsrmu0}
\end{eqnarray}
Actually, several terms of this expansion vanish as it will be seen below.

\subsection{Ground-state wave function for $\mu \to 0$}

Using the expansion of the interaction $v_{ee}^{\mu}$ (Eq.~\ref{veemclaurin}) and taking the functional derivatives of the expansions of the short-range Hartree, exchange and correlation energies (Eqs.~\ref{Usrmu0},~\ref{Exsrmu0} and~\ref{Ecsrmu0}) to obtain the expansion of the potential $v^{\mu}$, we arrive to the following formal expansion for the Hamiltonian of the fictitious system
\begin{eqnarray}
\hat{H}^{\mu} &=& \hat{H}_{KS} + \sum_{n=0}^{\infty} v_{ee}^{(2n+1)} \mu^{2n+1} + \sum_{n=0}^{\infty} v_{hx}^{(2n+1)} \mu^{2n+1} 
\nonumber\\
 && + \sum_{n=1}^{\infty} \sum_{k=1}^{\infty} v_{c}^{(2n+1,k)} \mu^{2n+1+k}.
\label{Hmu0}
\end{eqnarray}
In this equation, $\hat{H}_{KS}$ is the KS Hamiltonian, $v_{ee}^{(2n+1)}$ refers to the coefficients of expansion~(\ref{veemclaurin}), $v_{hx}^{(2n+1)}$ refers to the Hartree and exchange contributions to the coefficients of the expansion of $v^{\mu}$ and $v_{c}^{(2n+1,k)}$ refers to the correlation contribution. The two superscripts in $v_{c}^{(2n+1,k)}$ reflect the fact that the terms in the expansion of the correlation energy (Eq.~\ref{Ecsrmu0}) come from two sources: the interaction $v_{ee}^{\mu}$ and the pair density $n_{2,c}^{\mu}$. The ground-state wave function $\Psimu$ of this Hamiltonian can also be expanded with respect to $\mu$
\begin{eqnarray}
\Psimu = \Phi + \sum_{k=1}^{\infty} \Psi^{(k)} \mu^{k},
\end{eqnarray}
where $\Phi$ is the KS determinant. We now show that several terms in the expansion of $\Psimu$ actually vanish.

For the erf interaction, $v_{ee}^{(1)}$ and $v_{hx}^{(1)}$ are constants (cf. Eqs.~\ref{veemclaurin},~\ref{Usrmu0} and~\ref{Exsrmu0}), and consequently $v_{c}^{(1,k)}=0$ for all $k \ge 1$, thus $\Psi^{(1)} = 0$. As there is no term in $\mu^2$ in~(\ref{Hmu0}), we also have $\Psi^{(2)} = 0$. It implies in turn that in Eq.~(\ref{Pcseriesmu}) the terms $k \le 2$ vanish, and thus by Eq.~(\ref{Ecsrmu0}) $v_{c}^{(2n+1,k)}=0$ for $k \le 2$ and all $n \ge 1$. In particular, $v_{c}^{(3,1)}=v_{c}^{(3,2)}=0$ which leads to the expansion of the Hamiltonian
\begin{eqnarray}
\hat{H}^{\mu}_{\erf} &=& \hat{H}_{KS} + \left(v_{ee}^{(3)}+v_{hx}^{(3)}\right) \mu^3 
 + \left(v_{ee}^{(5)}+v_{hx}^{(5)}\right) \mu^5 
\nonumber\\
&&+ v_c^{(3,3)} \mu^6 + \cdots.
\end{eqnarray}
We therefore have the following expansion for the wave function
\begin{equation}
\label{psimuerf}
\Psimu_{\erf} = \Phi + \mu^3 \Psi^{(3)} + \mu^5 \Psi^{(5)} + \mu^6 \Psi^{(6)} + \cdots.
\end{equation}
Therefore, the terms $k=1,2,4$ in Eqs.~(\ref{Pcseriesmu}) and~(\ref{Ecsrmu0}) vanish for for the erf interaction.

For the erfgau interaction, $v_{ee}^{(1)}=v_{ee}^{(3)}=0$, and consequently $v_{hx}^{(1)}=v_{hx}^{(3)}=0$ and $v_{c}^{(3,k)}=0$ for all $k \ge 1$, thus $\Psi^{(1)} = \Psi^{(2)} = \Psi^{(3)}= \Psi^{(4)}=0$. It implies in turn that in Eq.~(\ref{Pcseriesmu}) the terms $k \le 4$ vanish, and thus by Eq.~(\ref{Ecsrmu0}) $v_{c}^{(2n+1,k)}=0$ for $k \le 4$ and all $n \ge 1$. In particular, $v_{c}^{(5,1)}=v_{c}^{(5,2)}=v_{c}^{(5,3)}=v_{c}^{(5,4)}=v_{c}^{(6,1)}=v_{c}^{(6,2)}=v_{c}^{(6,3)}=v_{c}^{(6,4)}=v_{c}^{(7,1)}=v_{c}^{(7,2)}=v_{c}^{(7,3)}=v_{c}^{(8,1)}=v_{c}^{(8,2)}=v_{c}^{(9,1)}=0$ which leads to the expansion of the Hamiltonian
\begin{eqnarray}
\hat{H}^{\mu}_{\erfgau} &=& \hat{H}_{KS} + \left(v_{ee}^{(5)}+v_{hx}^{(5)}\right) \mu^5 
 + \left(v_{ee}^{(7)}+v_{hx}^{(7)}\right) \mu^5 
\nonumber\\
&& + \left(v_{ee}^{(9)}+v_{hx}^{(9)}\right) \mu^9 
+ v_c^{(5,5)} \mu^{10} + \cdots.
\end{eqnarray}
We therefore have the following expansion for the wave function
\begin{eqnarray}
\Psimu_{\erfgau} &=& \Phi + \mu^5 \Psi^{(5)} + \mu^7 \Psi^{(7)} + \mu^9 \Psi^{(9)}
\nonumber\\
 & & + \mu^{10} \Psi^{(10)} + \cdots.
\label{psimuerfgau}
\end{eqnarray}
Therefore, the terms $k=1,2,3,4,6,8$ in Eqs.~(\ref{Pcseriesmu}) and~(\ref{Ecsrmu0}) vanish for the erfgau interaction.

\section{Adiabatic connection near the physical system}
\label{app:asymptoticexpansion}

In this appendix, we study the erf and erfgau adiabatic connections near the physical system, i.e. for large interaction parameter $\mu$.

\subsection{Short-range interaction for $\mu \to \infty$}
We start by deriving a distributional asymptotic expansion for large $\mu$ of the short-range electron-electron interaction $\bar{v}_{ee}^{\mu}(r)=1/r - v_{ee}^{\mu}(r)$ for the erf and erfgau interactions,
\begin{equation}
\bar{v}_{ee,\erf}^{\mu}(r)=\frac{\erfc(\mu r)}{r},
\label{veeerfc}
\end{equation}
\begin{equation}
\bar{v}_{ee,\erfgau}^{\mu}(r)=\frac{\erfc(\mu r)}{r} + \frac{2\mu}{\sqrt{\pi}} e^{-\frac{1}{3}\mu^2 r^2}.
\label{veeerfgauc}
\end{equation}
Let $f:\mathbb{R}^+ \to \mathbb{R}$ be a test function (i.e., of bounded support and infinitely differentiable) and consider the following integral
\begin{equation}
I=\int f(r) \bar{v}_{ee}^{\mu}(r) d\b{r}.
\label{Int}
\end{equation}
$f$ can be expanded into its Maclaurin series
\begin{equation}
f(r) = \sum_{n=0}^{m} \frac{f^{(n)}(0)}{n!} r^n + R_m(r),
\label{fmaclaurin}
\end{equation}
where the Lagrange remainder $R_m(r)$ is
\begin{equation}
R_m(r)=\frac{f^{(m+1)}(\theta r)}{(m+1)!} r^{m+1},
\end{equation}
with $0 \leq \theta \leq 1$. Inserting the first term of the right-hand-side of expansion~(\ref{fmaclaurin}) into~(\ref{Int}) gives the first contribution to integral $I$
\begin{equation}
I_1=\sum_{n=0}^{m} \frac{f^{(n)}(0)}{n!} \int r^n \bar{v}_{ee}^{\mu}(r) d\b{r},
\label{Int1}
\end{equation}
where the last integral can be easily evaluated
\begin{equation}
\int r^n \bar{v}_{ee}^{\mu}(r) d\b{r} = 4 \pi \int r^{n+2} \bar{v}_{ee}^{\mu}(r) d r = \frac{4 \sqrt{\pi}}{(n+2) \, \mu^{n+2}} A_n,
\label{intrnvee}
\end{equation}
with $A_{n,\erf}=\Gamma(\frac{n+3}{2})$ for the erf interaction and $A_{n,\erfgau}= \Gamma(\frac{n+3}{2}) + 3^{\frac{n+3}{2}} (n+2) \Gamma(\frac{n+3}{2})$ for the erfgau interaction. Thus, Eq.~(\ref{Int1}) becomes
\begin{equation}
I_1 = 4 \sqrt{\pi} \sum_{n=0}^{m} \frac{ A_n f^{(n)}(0) }{n! (n+2) \, \mu^{n+2}}.
\end{equation}
The second contribution to $I$ coming from the remainder writes
\begin{equation}
I_2=\int \frac{f^{(m+1)}(\theta r)}{(m+1)!}  r^{m+1} \bar{v}_{ee}^{\mu}(r) d\b{r}.
\end{equation}
We shall assume in addition that, for any $m$, the $(m+1)$-th derivative of $f$ is bounded, i.e. $|f^{(m+1)}(r)| \leq M_m$, then
\begin{eqnarray}
|I_2| &\leq& \frac{M_m}{(m+1)!} \int r^{m+1} \bar{v}_{ee}^{\mu}(r) d\b{r} 
\nonumber\\
      &=&  \frac{4 \sqrt{\pi} M_m A_{m+1}}{(m+1)! (m+3) \, \mu^{m+3}},
\end{eqnarray}
meaning that $I_2={\cal O}(1/\mu^{m+3})$. Finally, using the definition of the $n$-th derivative of the three-dimensional Dirac delta distribution $\delta^{(n)}(\b{r})$,
\begin{equation}
\int f(r) \delta^{(n)}(\b{r}) d\b{r} = (-1)^n f^{(n)}(0),
\end{equation}
we obtain the following distributional asymptotic expansion of $\bar{v}_{ee}^{\mu}(r)$ when $\mu \to \infty$
\begin{equation}
\bar{v}_{ee}^{\mu}(r) =  4 \sqrt{\pi} \sum_{n=0}^{m} \frac{ (-1)^n A_n}{n! (n+2) \, \mu^{n+2}} \delta^{(n)}(\b{r}) + {\cal O}(\frac{1}{\mu^{m+3}}).
\label{veemuinf}
\end{equation}
Note that if we apply a scale factor to the interaction parameter of the erfgau interaction $\mu \to c \mu$ such that $c=\sqrt(A_{n,\erfgau}/A_{n,\erf})=\left( 1+6\sqrt{3}\right)^{1/2} \approx 3.375$, then the asymptotic expansions of the erf and erfgau interactions are the same to leading order. This provides a criteria for comparison of the two interactions.

Following the same procedure on the derivative of the modified interaction with respect to $\mu$ which writes
\begin{equation}
\label{dveeerfdmu}
\frac{\partial \bar{v}_{ee,\erf}^{\mu}(r)}{\partial \mu} = - \frac{2}{\sqrt{\pi}} e^{-\mu^2 r^2},
\end{equation}
for the erf interaction, and
\begin{eqnarray}
\frac{\partial \bar{v}_{ee,\erfgau}^{\mu}(r)}{\partial \mu} &=& -\frac{2}{\sqrt{\pi}} e^{-\mu^2 r^2} + \frac{2}{\sqrt{\pi}} e^{-\frac{1}{3}\mu^2 r^2}
\nonumber\\
& & - \frac{4 \mu^2 r^2}{3 \sqrt{\pi}} e^{-\frac{1}{3}\mu^2 r^2},
\label{dveeerfgaudmu}
\end{eqnarray}
for the erfgau interaction, leads to a similar asymptotic expansion
\begin{equation}
\frac{\partial \bar{v}_{ee}^{\mu}(r)}{\partial \mu} =  - 4 \sqrt{\pi} \sum_{n=0}^{m} \frac{ (-1)^n A_n}{n! \, \mu^{n+3}} \delta^{(n)}(\b{r}) + {\cal O}(\frac{1}{\mu^{m+4}}).
\label{dveemudmuinf}
\end{equation}
which is just the derivative of expansion~(\ref{veemuinf}). In the following, we will apply Eqs.~(\ref{veemuinf}) and~(\ref{dveemudmuinf}) assuming that the corresponding test function $f(r)$ related to the pair density satisfies all the required assumptions which are reasonable for finite systems.

\subsection{Short-range Hartree energy for $\mu \to \infty$}
Introducing the trivial variable transformation $\b{r}_2 \to \b{r}_{12}$ and the spherical average of the density $n(\b{r}_{12})$
\begin{equation}
\tilde{n}(r_{12})=\frac{1}{4\pi} \int n(\b{r}_{12}) d\Omega_{\b{r}_{12}},
\end{equation}
the short-range Hartree energy writes
\begin{equation}
\bar{U}^{\mu} = \frac{1}{2} \iint n(\b{r}_1)\tilde{n}(r_{12}) \bar{v}_{ee}^{\mu}(r_{12}) d\b{r}_1 d\b{r}_{12}.
\end{equation}
Using then the distributional asymptotic expansion when $\mu \to \infty$ of the short-range interaction (Eq.~\ref{veemuinf}) and noting that $\tilde{n}(r_{12})$ can be expanded as an even series of $r_{12}$ around $r_{12}=0$, we obtain the asymptotic series of the short-range Hartree energy
\begin{eqnarray}
\bar{U}^{\mu} &=& 2\sqrt{\pi} \sum_{n=0}^{m} \frac{A_{2n}}{(2n)! (2n+2) \mu^{2n+2}}  
\nonumber\\
&& \times \int n(\b{r}) n^{(2n)}(\b{r})  d\b{r} + {\cal O}(\frac{1}{\mu^{2m+3}}),
\label{Usrmuinf}
\end{eqnarray}
where the notation $n^{(2n)}(\b{r})=(\partial^{2n} \tilde{n}(r_{12}) /\partial r_{12}^{2n})_{r_{12}=0} $ has been used for the density and its spherical-averaged derivatives.

\subsection{Short-range exchange energy for $\mu \to \infty$}

Similarly, the short-range exchange energy can be written as
\begin{equation}
\bar{E}_{x}^{\mu} = \frac{1}{2} \iint \tilde{n}_{2,x}(\b{r}_1,r_{12}) \bar{v}_{ee}^{\mu}(r_{12}) d\b{r}_1 d\b{r}_{12},
\end{equation}
where $\tilde{n}_{2,x}(\b{r}_1,r_{12})$ is the spherical-average exchange pair density which can be expanded like the density as an even series of $r_{12}$ around $r_{12}=0$ (``no cusp for exchange''). We therefore obtain the asymptotic series of the short-range exchange energy
\begin{eqnarray}
\bar{E}_{x}^{\mu} &=& 2\sqrt{\pi} \sum_{n=0}^{m} \frac{A_{2n}}{(2n)! (2n+2) \mu^{2n+2}}  
\nonumber\\
&& \times \int n_{2,x}^{(2n)}(\b{r},\b{r})  d\b{r} + {\cal O}(\frac{1}{\mu^{2m+3}}),
\label{Exsrmuinf}
\end{eqnarray}
with the on-top exchange pair density and its spherical-averaged derivatives $n_{2,x}^{(2n)}(\b{r},\b{r})=(\partial^{2n} \tilde{n}_{2,x}(\b{r},r_{12}) /\partial r_{12}^{2n})_{r_{12}=0} $.

\subsection{Short-range correlation energy for $\mu \to \infty$}

The asymptotic expansion of the short-range correlation energy when $\mu \to \infty$ can formally be found by considering its derivative (cf. Eq.~\ref{dEcdmu})
\begin{equation}
\label{dEcdmusphavg}
\frac{\partial \bar{E}_{c}^{\mu}}{\partial \mu} = \frac{1}{2} \iint \tilde{n}_{2,c}^{\mu}(\b{r}_1,r_{12}) \frac{\partial \bar{v}_{ee}^{\mu}(r_{12})}{\partial \mu} d\b{r}_1 d\b{r}_{12},
\end{equation}
where $\tilde{n}_{2,c}^{\mu}(\b{r}_1,r_{12})$ is the spherical average of $n_{2,c}^{\mu}(\b{r}_1,\b{r}_2)$, inserting the asymptotic expansion of $\tilde{n}_{2,c}^{\mu}(\b{r}_1,r_{12})$ and the distributional asymptotic expansion of $\partial \bar{v}_{ee}^{\mu}/\partial \mu$ (Eq.~\ref{dveemudmuinf}) and re-integrating with respect to $\mu$. Actually, considering only the first term of the asymptotic expansion of the correlation pair
density, i.e. $\tilde{n}_{2,c}^{\mu}(\b{r}_1,r_{12}) = \tilde{n}_{2,c}(\b{r}_1,r_{12}) +\cdots$, is sufficient to find the first term of the asymptotic expansion  of the correlation energy. Similarly to the exchange energy, the leading term of the short-range correlation energy for large $\mu$ is consequently given by the on-top correlation pair density $n_{2,c}(\b{r},\b{r})$
\begin{equation}
\bar{E}_{c}^{\mu} = \frac{\sqrt{\pi} A_0 }{\mu^2} \int n_{2,c}(\b{r},\b{r})
d\b{r} + \cdots.
\label{Ecsrmuinf1}
\end{equation}
Notice that, for the erf interaction, this result has already been derived~\cite{PolColLeiStoWerSav-IJQC-03}.

Using Eqs.~(\ref{Usrmuinf}), (\ref{Exsrmuinf}) and~(\ref{Ecsrmuinf1}), the external local potential $v^{\mu}(\b{r})= v_{ne}(\b{r}) + \delta \bar{U}^{\mu}/ \delta n(\b{r}) + \delta \bar{E}_{xc}^{\mu}/ \delta n(\b{r})$ has the following behavior for large $\mu$
\begin{equation}
{v}^{\mu}(\b{r}) = v_{ne}(\b{r}) + \frac{\sqrt{\pi} A_0}{\mu^2} \frac{\delta}{\delta n(\b{r})} \int  n_{2}(\b{r}',\b{r}') d\b{r}' + \cdots,
\end{equation}
where $n_{2}(\b{r},\b{r})$ is the total on-top pair density. As $v_{ee}^{\mu}$ has a similar expansion for $\mu \to \infty$ beginning with $\mu^{-2}$ (see Eq.~\ref{veemuinf})
\begin{equation}
v_{ee}^{\mu}(r) = \frac{1}{r} - \frac{2 \sqrt{\pi} A_0 }{\mu^2}  \delta(\b{r}) + \cdots.
\end{equation}
the behavior for large $\mu$ of the modified Hamiltonian $\hat{H}^{\mu} = \hat{T} +\sum_{i<j} {v}_{ee}^{\mu}(r_{ij}) + \sum_i v^{\mu}(\b{r}_i)$ is of the form
\begin{equation}
\hat{H}^{\mu} = \hat{H} + \frac{1}{\mu^2} \hat{H}^{(2)} +\cdots,
\end{equation}
The corresponding wave function $\Psimu$ and consequently the correlation pair density $\tilde{n}_{2,c}^{\mu}(\b{r}_1,r_{12})$ has a similar asymptotic expansion
\begin{equation}
\label{Pcmuinf}
n_{2,c}^{\mu}(\b{r}_1,r_{12}) = n_{2,c}(\b{r}_1,r_{12}) + \frac{1}{\mu^2} n_{2,c}^{(2)}(\b{r}_1,r_{12}) + \cdots.
\end{equation}
The absence of term in $1/\mu$ in the last expansion implies that the first two terms in the asymptotic expansion of $\partial \bar{E}_{c}^{\mu} / \partial \mu$ can be determined by considering only $n_{2,c}(\b{r}_1,r_{12})$. We find
\begin{eqnarray}
\frac{\partial \bar{E}_{c}^{\mu}}{\partial \mu} &=& - \frac{2 \sqrt{\pi} A_0 }{\mu^3} \int n_{2,c}(\b{r},\b{r}) d\b{r} 
\nonumber\\
&& - \frac{2 \sqrt{\pi} A_1}{\mu^4} \int n_{2,c}'(\b{r},\b{r}) d\b{r} +\cdots,
\end{eqnarray}
where $n_{2,c}'(\b{r},\b{r})$ is the spherical-averaged derivative of the on-top correlation pair density which, according to the electron-electron cusp condition~\cite{Kim-PRA-73}, is equal to the total on-top pair density: $n_{2,c}'(\b{r},\b{r})=n_{2}(\b{r},\b{r})$. Therefore, for large $\mu$ the short-range correlation energy has the exact behavior
\begin{equation}
\label{Ecsrmuinf2}
\bar{E}_{c}^{\mu} = \frac{\sqrt{\pi} A_0}{\mu^2} \int n_{2,c}(\b{r},\b{r}) d\b{r} + \frac{2 \sqrt{\pi} A_1}{3 \mu^3} \int n_{2}(\b{r},\b{r}) d\b{r} +\cdots.
\end{equation}

\bibliographystyle{apsrev}
\bibliography{biblio}

\end{document}